\documentclass[9pt, sigconf]{acmart}

\AtBeginDocument{%
  }

\usepackage{graphicx}
\usepackage{balance}
\usepackage{tikz}

\usepackage{enumitem}
\usepackage[utf8]{inputenc}
\usepackage{algorithm}
\usepackage[noend]{algpseudocode}
\usepackage{xcolor, soul}
\usepackage{cleveref}
\usepackage{arydshln}

\usepackage{booktabs}

\newcommand{\Autoref}[1]{%
  \begingroup%
  \def\chapterautorefname{Chapter}%
  \def\sectionautorefname{Section}%
  \def\subsectionautorefname{Subsection}%
  \def\subsubsectionautorefname{Subsubsection}%
  \autoref{#1}%
  \endgroup%
}
\copyrightyear{2024}
\acmYear{2024}
\setcopyright{rightsretained}
\acmConference[SENSYS '24]{The 22nd ACM Conference on Embedded Networked Sensor Systems}{November 4--7, 2024}{Hangzhou, China}
\acmBooktitle{The 22nd ACM Conference on Embedded Networked Sensor Systems (SENSYS '24), November 4--7, 2024, Hangzhou, China}
\acmDOI{10.1145/3666025.3699359}
\acmISBN{979-8-4007-0697-4/24/11}

\acmSubmissionID{491}

\begin{document}

% \title{Hawk: An Accurate and Efficient Non-intrusive Appliance Load Monitoring System}
\title{Hawk: An Efficient NALM System for Accurate Low-Power Appliance Recognition}
% \affiliation{%

\author{Zijian Wang$^{1, 2}$, Xingzhou Zhang$^{1, 2}$, Yifan Wang$^{1, 2}$, Xiaohui Peng$^{1, 2}$, Zhiwei Xu$^{1, 2, 3}$}
\affiliation{
 \institution{$^1$Institute of Computing Technology, Chinese Academy of Sciences, Beijing, China}
 \institution{$^2$University of Chinese Academy of Sciences, Beijing, China}  
 \institution{$^3$Great Bay University, Dongguan, China}
 \country{China}
}
\email{{wangzijian19z, zhangxingzhou, wangyifan2014, pengxiaohui, zxu}@ict.ac.cn}

% \author{Zijian Wang}
% \affiliation{%
%   \institution{Institute of Computing Technology, Chinese Academy of Sciences}
%   \institution{University of Chinese Academy of Sciences}
%   % \streetaddress{1 Th{\o}rv{\"a}ld Circle}
%   \city{Beijing}
% \email{wangzijian19z@ict.ac.cn}

% \author{Xingzhou Zhang}
% \affiliation{%
%   \institution{Institute of Computing Technology, Chinese Academy of Sciences}
%   \institution{University of Chinese Academy of Sciences}
%   \city{Beijing}
%   \country{China}
% }
% \email{zhangxingzhou@ict.ac.cn}

% \author{Yifan Wang}
% \affiliation{%
%   \institution{Institute of Computing Technology, Chinese Academy of Sciences}
%   \city{Beijing}
%   \country{China}
% }
% \email{wangyifan2014@ict.ac.cn}

% \author{Xiaohui Peng}
% \affiliation{%
%   \institution{Institute of Computing Technology, Chinese Academy of Sciences}
%   \institution{University of Chinese Academy of Sciences}
%   \city{Beijing}
%   \country{China}
% }
% \email{pengxiaohui@ict.ac.cn}

% \author{Zhiwei Xu}
% \affiliation{%
%   \institution{Institute of Computing Technology, Chinese Academy of Sciences}
%   \institution{University of Chinese Academy of Sciences}
%   \institution{Great Bay University}
%   \city{Beijing}
%   \country{China}
% }
% \email{zxu@ict.ac.cn}

\renewcommand{\shortauthors}{Z. Wang et al.}
\renewcommand{\authors}{Zijian Wang, Xingzhou Zhang, Yifan Wang, Xiaohui Peng, Zhiwei Xu}

\begin{abstract}

Non-intrusive Appliance Load Monitoring (NALM) aims to recognize individual appliance usage from the main meter without indoor sensors. However, existing systems struggle to balance dataset construction efficiency and event/state recognition accuracy, especially for low-power appliance recognition.
This paper introduces Hawk, an efficient and accurate NALM system that operates in two stages: dataset construction and event recognition. In the data construction stage, we efficiently collect a balanced and diverse dataset, HawkDATA, based on balanced Gray code and enable automatic data annotations via a sampling synchronization strategy called shared perceptible time. 
During the event recognition stage, our algorithm integrates steady-state differential pre-processing and voting-based post-processing for accurate event recognition from the aggregate current. 
Experimental results show that HawkDATA takes only 1/71.5 of the collection time to collect 6.34x more appliance state combinations than the baseline.
In HawkDATA and a widely used dataset, Hawk achieves an average F1 score of 93.94\% for state recognition and 97.07\% for event recognition, which is a 47. 98\% and 11. 57\% increase over SOTA algorithms.
Furthermore, selected appliance subsets and the model trained from HawkDATA are deployed in two real-world scenarios with many unknown background appliances. The average F1 scores of event recognition are 96.02\% and 94.76\%.
Hawk's source code and HawkDATA are accessible at
% \url{https://anonymous.4open.science/r/Hawk4NALM2/}.
\url{https://github.com/WZiJ/SenSys24-Hawk}.

\end{abstract}




% \received{15 November 2023}

\maketitle

\section{INTRODUCTION} \label{introduction}
Indoor human activity recognition (HAR) is an attractive issue. Appliance usage, as an essential part of indoor activities, reflects people's daily routines and impacts household energy consumption. Accurate detection of appliance usage enables various applications, such as detecting hazardous appliances (e.g. chargers for flammable lithium batteries), elderly care\cite{suryadevara2013forecasting}, building energy management\cite{wu2019concatenate, li2015smart}, and optimizing operating plans for residential\cite{valovage2018model} and industrial users\cite{luan2022industrial}. Non-intrusive appliance load monitoring (NALM)\cite{hart1985nilmproto} detects the usage of individual appliances by monitoring the main circuit without indoor sensors. Due to its ease of deployment, privacy protection, and low hardware cost, NALM is very attractive\cite{sun2020vibrosense,patel2007flick} and is considered vital for smart meters\cite{thokala2022effective, gupta2023enabling}.

Despite its potential, existing NALM systems struggle to balance algorithm accuracy with the efficiency of dataset construction. NALM is a typical single-channel blind source separation (SCBSS) problem\cite{zhang2018sequence}, which is highly underdetermined\cite{he2023msdc} and challenging. Current algorithms mainly focus on recognizing a few noticeable appliances\cite{zhang2018sequence, he2023msdc, bejarano2019deep} and are hard to identify low-power ones from the total current. 
Moreover, long-term data collection is necessary to capture diverse real-world samples for model training\cite{kelly2015uk, wang2024sudokusens}, and manual inspection is essential for accurate data annotation\cite{filip2011blued}, which becomes a burden for supervised NALM algorithms\cite{zhu2023combining}. Therefore, building a NALM system that accurately identifies all household appliances, especially low-power ones, faces three challenges:

\textbf{Inefficiency of dataset collection.} Like most HAR scenarios, NALM data collection suffers from \textit{data scarcity}\cite{wang2024sudokusens} and \textit{imbalance}\cite{zhang2019new, liu2022open}. Most appliances, such as 86\% in the BLUED dataset\cite{filip2011blued}, are switched less than ten or even zero times a day. In addition, the distribution of the usage of different appliances is highly imbalanced. The distribution of usage varies over time and in different households, making it difficult for the training sets collected to represent various real-world conditions\cite{wang2024sudokusens}. Therefore, traditional NALM data collection involves long-term data acquisition in different households\cite{kelly2015uk}, which is time consuming and error prone\cite{chen2021sensecollect}. Imbalanced appliance usage leads to redundant samples\cite{zhang2019new, tarekegn2021review, wang2021lcl} and insufficient usage data for some appliances\cite{filip2011blued, yu2024multi}, which burdens feature extraction and model training\cite{tarekegn2021review}.

\textbf{Inefficiency of dataset annotation.} Traditional NALM systems collect the aggregated current and individual appliance status for data labeling during the dataset construction phase. Sampling synchronization in such typical distributed acquisition scenarios affects the accuracy of the labeling\cite{renaux2020lit, filip2011blued}. Traditional synchronization methods\cite{kriechbaumer2018blond, renaux2020lit} are two-step processes: synchronizing the sensor nodes' clocks to a standard clock, followed by responding and timestamping the sensor data. This \textit{ two-step sync} introduces high upper-bound cumulative errors. Thus, manual inspection\cite{filip2011blued} is essential for accurate event timestamps, which is time-consuming and error-prone, especially for low-power appliances.

\textbf{Inaccuracy of the low-power appliances recognition}: NALM is a typical SCBSS task that separates individual source signals from a single aggregated signal (typically total current or total power). The identification of target appliances is affected by other appliance currents and electrical fluctuations. Inspired by the Signal-to-Interference-plus-Noise Ratio(SINR) for assessing the separability of SCBSS\cite{zhang2022quantifying}, lower-power appliances are generally more difficult to recognize with the same level of interference. Thus, many SOTA algorithms\cite{he2023msdc,zhang2018sequence} only identify a few apparent appliances and have a high computational overhead\cite{wang2021calm}, which become key obstacles for \textit{using one sensor for all household appliances}. 

\begin{figure}
  \begin{minipage} {\linewidth}
    \centering
    \includegraphics[width=0.9\linewidth]{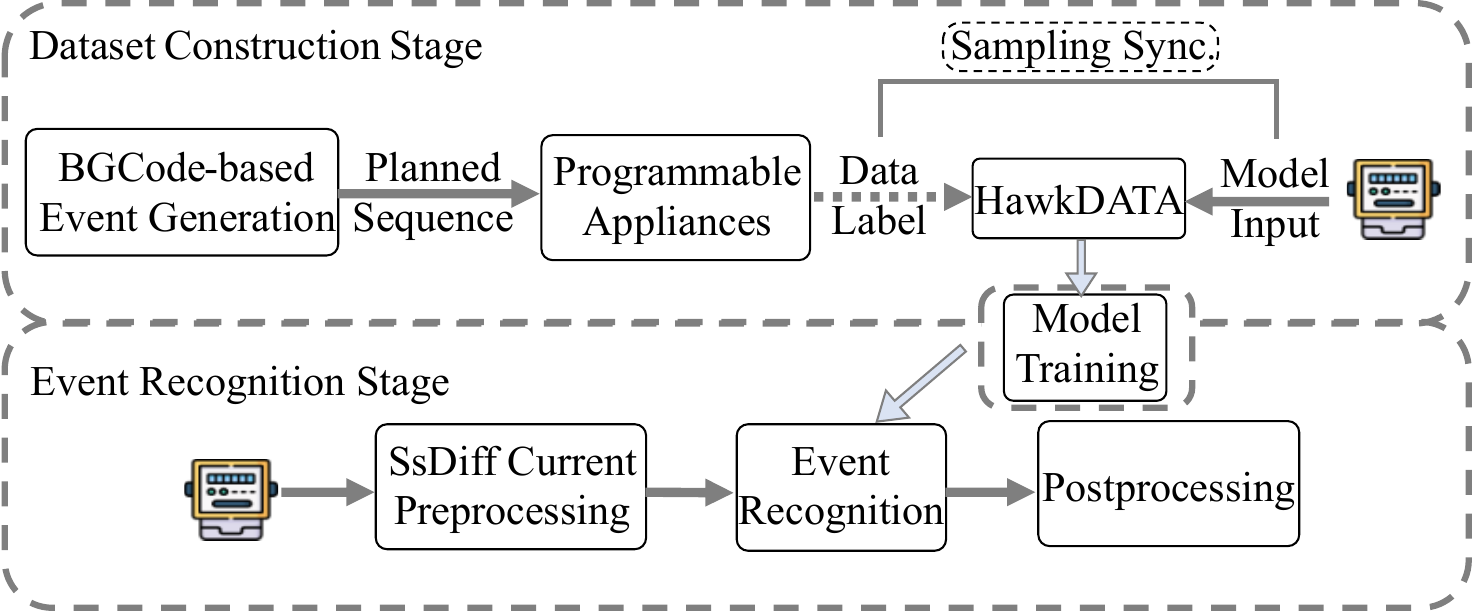}
    \caption{The methodological pipeline of Hawk.}
    \label{fig:high-frequency-NIALM-pipeline}
  \end{minipage}
\end{figure}

To summarize the above challenges, we pose the following question: \textit{can we \textbf{efficiently} train a model capable of \textbf{accurately} recognizing (F1 score > 90\%) \textbf{low-power}(< 50w) appliances from aggregate current in \textbf{real world} with low computational overhead?} To this end, we propose Hawk, an efficient NALM system for accurate low-power appliance recognition. As illustrated in \autoref{fig:high-frequency-NIALM-pipeline}, Hawk's methodological pipeline is divided into two stages: dataset construction and event recognition. First, Hawk generates diverse and balanced event sequences using \textit{grouped randomized balanced Gray code (BGCode)}. Programmable appliances execute schedules while their current is recorded to label the aggregated current. Then, high-frequency sampled data from distributed sensors are synchronized using a \textit{shared perceptible time} strategy, which eliminates the intermediate clock in the two-stage sync process. During the event recognition stage, a \textit{steady-state differential (SsDiff) current} pre-processing method enhances the SINR of appliances events, especially low-power ones, from the aggregated current.The post-processing stage reports events after a popularity-based voting method to improve the algorithm's robustness to false predictions.

The application scenario of the Hawk system is also illustrated in \autoref{fig:high-frequency-NIALM-pipeline}. Model developers efficiently construct datasets with a set of appliances and train the corresponding models to recognize these appliances of the same models. Users may be interested in any subset of the collected appliances and download the corresponding models to recognize the usage of the appliances against unknown background appliances.
Identifying appliances by appliance model imposes stronger constraints than recognition by type, but it is more reasonable. This is because the electric signature of an appliance is primarily determined by its front-end circuitry\cite{he2012front} and power consumption, rather than by a strong mapping to the type of appliance. Thus, two appliances of the same type may have different electrical signatures, and two appliances with identical electrical signatures may be of entirely different types. Consequently, previous studies that validate the recognition accuracy in the same appliance type in different households\cite{kelly2015uk} always result in a low recognition accuracy\cite{zhang2018sequence, bejarano2019deep}, which complicates the evaluation of the effectiveness of the algorithm\cite{he2023msdc}.

To our knowledge, Hawk is the first NALM system to accurately recognize low-power appliances from aggregated current using the model trained in the laboratory and inference in the wild. Full-stack optimizations of the Hawk system contribute to the final recognition accuracy.
The Hawk sampling synchronization strategy (like the coordinated claws) reduces the maximum error bound to a sampling interval, 1/17.2 of 802.11 TSF. It enables automated data annotation to reduce labor costs and potential deviation.
HawkDATA (like huge and balanced wings) achieves 6.34 times more appliance state combinations in just 1/71.5 of collection time compared to baseline\cite{pereira2022sustdataed2}. It also improves the category balance ratio of the event and state combination by 151.4 and 7188.5 times. Hawk's algorithm (like the sharp eyes) achieves an average F1 score of 92.71\% for event recognition and 93.93\% for state identification in HawkDATA. The F1 score of state identification increases by 47.98\% compared to \cite{he2023msdc, wang2021calm}. With the same model trained on HawkDATA, Hawk achieves an average F1 score of 96.03\% with five appliances in a residential area and 94.76\% with six appliances, including five low-power ones, in an official area, both with many unknown backgrounds appliances. In the BLUED dataset\cite{filip2011blued}, Hawk's average F1 score for event classification reaches 97.07\%, improved by 11.57\% to SOTA\cite{yu2024multi}. In addition, Hawk's average cost of single-point, high-frequency (16Khz, 24-bit) stabilized sampling is reduced to \$4.12, about 1/3 of Hz-level sampling board\cite{debruin2015powerblade}.

The rest of the paper is organized as follows: \Autoref{background} introduces related work. \Autoref{development} and \Autoref{deployment} detail dataset construction scheme and algorithm pipeline. \Autoref{UseCase} describes hardware prototype, evaluation settings and in-the-wild deployment. \Autoref{evaluation} presents full-stack evaluation results. \Autoref{discussion} discuss limitation and failure scenario of Hawk. \Autoref{conclusion} concludes the paper.

\section{RELATED WORK} \label{background}

\textbf{NALM Sampling Synchronization.}
The NALM datasets capture the current or power consumption in the main circuit as input to the model and collect the current or power consumption of the individual appliances for data annotation.
In such a typical distributed data sampling scenario, accurate clock synchronization is the foundation of accurate automated annotation. \cite{campbell2018energy} and \cite{viswanathan2016exploiting} rely on hardware voltage zero crossing detectors to synchronize sampling and clocks. However, such hardware-based methods are vulnerable to voltage fluctuation due to hardware error\cite{kelly2015uk} or some appliances events (e.g., some low-end fans). Most works adopt network-based synchronization methods. BLOND \cite{kriechbaumer2018blond} employs NTP with one deviation tested as 6.8 ms, while LIT2020 \cite{renaux2020lit} uses a wireless clock distributor based on a 433MHz RF transmitter with a precision of less than 5 ms. BLUED\cite{filip2011blued} uses a collision-free wireless protocol but still needs visual inspections to realign the annotation of events. Other fields include \cite{chen2021understanding} achieving 1us-accuracy PTP-like protocol in WiFi networks via TSF. However, the performance of such wireless methods varies with the network jitters\cite{thi2022ieee}.

\textbf{NALM Dataset Construction.}
Like many HAR scenarios, NALM suffers from data scarcity\cite{chen2021sensecollect}. Thus, traditional NALM dataset construction extend data collection periods and scenarios to capture diverse, authentic electricity usage behaviors. REDD\cite{kolter2011redd} and UK-DALE\cite{kelly2015uk} collect data from numerous residential houses over long periods. BLOND\cite{kriechbaumer2018blond} collects a high-frequency dataset in office settings. However, such long-period data collection is time-consuming and error-prone\cite{chen2021sensecollect}. What is worse, the scarce and imbalanced usage of appliances always results in an imbalanced dataset: redundant samples for certain state combinations\cite{zhang2019new, tarekegn2021review, wang2021lcl}, or absence of several appliances' activity\cite{filip2011blued, zhu2023combining}. Some datasets\cite{kahl2016whited, kahl2016whited} choose controlled laboratory conditions to automate the data collection process, which consists of many appliances. However, they lack modeling for balanced state combinations and only capture current from individual appliances, missing aggregate current in NALM settings. Moreover, it is challenging to generate reliable synthetic data\cite{wang2024sudokusens, renaux2020lit} since the appliance current is an analog variable with multiple influences and complex interference between appliances. 

\textbf{NALM State/Event Identification.}
Existing NALM systems try to identify the state or event of appliances from aggregated current or power. State identification algorithms recognize appliance statuses, basic operational states\cite{he2023msdc}, and power consumption\cite{zhang2018sequence}. Event recognition algorithms detect and classify state transitions\cite{hart1992nonintrusive, patel2007flick, wu2019concatenate}.
Previous studies have used features at various frequencies for NALM systems. 
\cite{hart1992nonintrusive} pioneers the use of \textbf{low-frequency} features through an unsupervised learning pipeline. HMMs and their variants are prevalent \cite{shaloudegi2016sdp, kolter2012approximate} and are used in industrial cases \cite{luan2022industrial}. Deep learning based on low-frequency features also demonstrates promise\cite{kelly2015neural, bejarano2019deep, zhang2018sequence, he2023msdc, yue2020bert4nilm, faustine2020unet}.
To address the diversity in dataset formats and algorithm design, 
NILMTK\cite{batra2019towards} provides a unified dataset parsing interface and integrates several classic low-frequency NALM algorithms for algorithm comparison. 
\cite{kalinke2021evaluation} validates NALM different algorithms in industrial settings.  
While low-frequency features require less computation and economic infrastructure, these approaches may struggle to detect low-power appliances, distinguish similar-power ones, and perform poorly in energy consumption trace disaggregation\cite{batra2015if}. Some works\cite{gupta2010electrisense, patel2007flick} adopt \textbf{ultra-high-frequency} noise sampling (over 1 MHz) to identify electrical events and recognize devices equipped with SMPS. However, noise is attenuated along the cable\cite{patel2007flick, wu2019concatenate}, limiting the distance between the sensor and appliances. Additionally, the hardware infrastructure required for ultra-high-frequency sampling is much more expensive, and a considerable computational overhead exists. Recent work \cite{wang2021calm, yu2024multi, wu2019concatenate} utilize \textbf{high-frequency} (1 kHz to 1 MHz) currents with DNNs and achieve promising results. \cite{gupta2023enabling} utilizes EMI to recognize server-level power consumption, even distinguishing between different servers of the same model.
% However, the types of appliances it could recognize are limited to several PFC circuits.

\section{Dataset construction stage} \label{development}
The NALM dataset consists of two stages: dataset collection and data annotation. The collection process can be simplified as collecting aggregated and individual signals while a set of appliances $A$ executes an events sequence $S$. 
Hawk makes the process efficient by automating the execution of balanced and diverse sequences using programmable appliances. Additionally, Hawk synchronizes the distributed sampling by shared perceptible time, enabling automated data annotation and significantly reducing labor costs.

\subsection{Grouped Randomized Balanced Gray Code} \label{GRBGCode}
Hawk aims to efficiently collect a NALM dataset that satisfies the balance and diversity demands of two primary tasks: event and state recognition. This goal comes from Hawk's application scenarios, where users are interested in recognizing any subset of $A$ with arbitrary background appliances. Therefore, there should be no assumed background currents and priority among appliances.

However, in traditional NALM dataset construction\cite{kelly2015uk, kolter2011redd}, appliance sets $A_{\text{real}}$ are chosen from a limited number of volunteer households, and event sequences $S_{\text{real}}$ are executed naturally by the household residents. The $S_{\text{real}}$ is naturally temporally sparse and biased according to the number of residents, their appliance usage habits, and exogenous conditions. This leads to long-term data collection periods for diverse but imbalanced samples, requiring additional software preprocessing over TB-level volumn of raw data, which burdens algorithm users. Thus, some works\cite{kahl2016whited, renaux2020lit, picon2016cooll} collect datasets in a controlled laboratory to bridge appliance event execution sequence with demands of model training. However, they lack an abstraction of appliance event execution to satisfy the diverse and balanced demands of event/state-based algorithm model training, and most only collect individual currents\cite{kahl2016whited, picon2016cooll}.

\begin{figure}
  \begin{minipage}[t]{\linewidth} 
    \centering
    \includegraphics[width=0.95\linewidth]{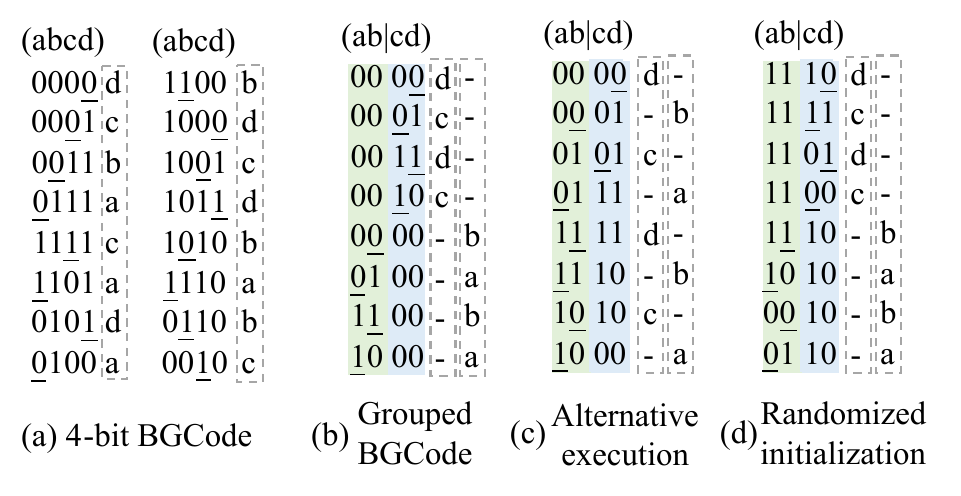}
    \caption{Four-bit balanced Gray code and our proposed variant for four appliances (a/b/c/d).
    % The 4-bit 0-1 sequence represents the switch states of four appliances, while the numbers enclosed in dashed boxes indicate the appliance IDs with events execution. 
    % The middle one starts from a randomized state but follows the same sequence. The right one starts from the same initial state but from a different start point(as the green array points) of the circular sequence.
    }
    \label{fig:BalancedGrayCode}
  \end{minipage}
\end{figure}

Therefore, we model appliance OFF-ON states as 0-1 and propose an event generation strategy based on a variant of balance Gray code (BGCode). As shown on the (a) of \autoref{fig:BalancedGrayCode}, the BGCode has the following properties that make it ideal for generating event sequence that guides the execution of appliance events.
\begin{itemize}
    \item Balance: BGCode ensures a nearly balanced flip count for each bit, balancing switch counts among appliances. 
    \item One-bit clipping: BGCode differs only one bit among adjacent states, aligning with independent appliance switches.
    \item No duplicate travel: BGCode travels the entire state space with no duplicate state, minimizing switching operation.
    \item Scalability: BGCode can be extended to present any even number of appliances' OFF-ON states\cite{robinson1981counting}.
\end{itemize}

Traversing the entire state space of $2^n$ combinations becomes extremely time-consuming when the number of appliances becomes large. Additionally, each state combination must stay for sufficient time intervals to collect adequate samples. Thus, we introduce grouping and randomization strategies in event generation to reduce the size of event sequences while maintaining a diverse and balanced property.
\textbf{Grouping} aims to reduce the scale of execution of balanced Gray code sequences. Given that we have n (4 in (a) of  \autoref{fig:BalancedGrayCode}) appliances divided evenly into m groups (2 in (b) of \autoref{fig:BalancedGrayCode}), each group is selected to execute complete balanced Gray code sequences internally. Then, the number of events required is reduced from $2^{n}$ to $m*2^{n/m}$. Such a sequence still meets the requirement of balance, but limits the range of traversal. Therefore, the grouped strategy will be executed in multiple rounds to achieve a more diverse combination of states. In each round, one or more groups of appliances will be activated and events within the groups will be alternated as (c) of \autoref{fig:BalancedGrayCode}. To reduce the probability of state combination collisions in multiple generation rounds, we introduce \textbf{Randomization} in terms of grouping, execution sequences and initial state within groups, as shown in (d) of \autoref{fig:BalancedGrayCode}. 

In addition, this strategy comes with certain constraints. For example, the number of groups activated per round is limited to ensure safe automated execution of electricity usage without supervision. Additionally, to maintain the balance of the overall data set, the activation frequency of each group must be maintained throughout the process. Finally, in the training set generation phase, Algorithm 1 is executed for 30 rounds, yielding a sequence of 2880 events and 2584 unique states. As for the validation set, we performed the strategy for 18 rounds, resulting in 2112 events and 1962 unique states. This data generation strategy ensures a low overlap between the training and test sets, less than 17\%, requiring the model to learn the underlying data patterns. The generated sequences are stored in a file and executed by the programmable appliances described in the following subsection.

% In the training set generation phase, the strategy is executed for 30 rounds, yielding 2,880 events and 2,584 unique states. As for the testing set, we perform the strategy for 18 rounds, resulting in 2112 events and 1962 unique states. The planned number of events executed for different appliances differs by no more than 2. During data collection, each state maintains the same duration to collect an equal number of 20ms period samples.
% This data generation strategy ensures a low overlap between the training and testing sets, less than 17\%, requiring the model to learn the underlying data patterns. Finally, we store the generated sequences in a file and execute them using the programming appliances. 

\subsection{Programmable Appliance} \label{ProApp}
In the actual dataset collection process, it is necessary to implement programmable appliances that execute generated event sequences automatically without ambiguity and maintain the consistency of the appliance's state in cyberspace and physical space. Hence, the following two requirements must be satisfied: accurate event execution and timely feedback of the appliance state.

The programmable appliance is a logical entity implemented with an event executor, an individual current sensor, and a controller. The event executors consist of esp32-controlled servos and relays to simulate human operations, and ESP32s provide interfaces as Restful servers. Each executor is assigned a static address and pre-calibrated to ensure \textbf{accurate event execution}. The controller detects the state from the cycle-level root mean squared (RMS) current collected by the individual current sensor. However, some appliances exhibit continual variations in the RMS current. Hence, we denoised the RMS current waveform by averaging values in a window of length $l$, specific for each appliance, excluding the $n$ largest and the $m$ smallest values. And we switch the appliances multiple times in advance to determine these parameters corresponding to specific appliances. We keep $l$ as short as possible (averaged 60 ms) to enable \textbf{timely feedback of the appliance status}. When the appliance is turned on or off due to a malfunction or timed switch, the abnormal event will be reported and automatically corrected.

%The dataset labeling is also based on the strategy. Additionally, manual inspection is essential to ensure the accuracy of dataset annotation. Yet, our inspection only needs to verify the value of the parameter settings for individual appliances from their current. Relying on the clock synchronization strategy called shared perceptible time, we can regard the labeling of individual appliances as accurate labeling of the corresponding total current and significantly improve the efficiency of data labeling.

\subsection{Shared Perceptible Time}
Clock synchronization accuracy is crucial for NALM data annotation, especially for datasets with larger granularity annotations, such as the 20ms-cycle level for HawkDATA and the 6s level for UK-DALE\cite{kelly2015uk}. Given a synchronization strategy with a maximum deviation of $E$, events occurring within $E$ around the label bounds may cause a label-level bias, which misguides model training and data analysis. Traditional NALM data acquisition suffer from the cumulative error of the \textit{two-step sync}. Clock synchronization protocols in cyberspace are always influenced by network jitters. Additionally, the delay in CPU/MCU data response can be unstable. For example, using FPGA for high-frequency ADC data buffering may cause fluctuations in event timestamping\cite{kriechbaumer2018blond}.

However, NALM dataset construction focuses more on internal synchronization than external synchronization. Hence, we propose a sampling synchronization strategy called shared perceptible time to synchronize sampling directly. The core of the strategy is that time can be represented not only as counters driven by crystal oscillators in cyberspace but also by continuously varying physical signals. In the context of NALM dataset construction, most appliances are connected in parallel and share the same voltage within a household. We can utilize multichannel simultaneous ADCs to capture current alongside the voltage and timestamp each sample with (cycle number, phase value) from the voltage value. Cycle ID is confirmed by the TSF value of the first sample in the voltage cycle. The phase value is assigned a continuous distance to the first zero-crossing point, which is quantified by the sampling interval.
We assume that the sampling interval remains stable over a short period, such as a voltage cycle of 20 ms, so the interval can be used to quantify the phase position. Ultimately, we construct the global current sequence by constructing the global voltage sequence.

\section{Event Recognition Stage} \label{deployment}
NALM algorithms suffer from low signal-to-interference-plus-noise ratios(SINR), where signals and fluctuations from other appliances influence targeted appliance recognition. Hawk's enhances the target events' SINR by steady-state differential current pre-process and improves the algorithm's robustness by a popularity-based voting post-processing method. Thus, some basic classifiers can achieve high accuracy and robustness in recognition while maintaining low computational overhead.

% One of the most fundamental challenges in NALM is accurately recognizing appliance states and events despite the complex combination of appliance states and noise in the background. During deployment, Hawk significantly enhances the SINR of appliance events to be identified using a pre-processing method called steady-state differential current. Additionally, a post-processing method based on voting within time windows is employed to reduce the impact of noise-induced false predictions on the reporting results. Finally, we combined these two strategies with different classifiers to construct Hawk's algorithm pipeline, achieving real-time, high-accuracy appliance event identification.

% This section mainly explains how Hawk uses steady-state differential current to enhance target appliances' Signal-to-Interference-plus-Noise Ratio (SINR), especially low-power ones', from aggregated current.

\subsection{Steady-state Differential Current}
Unlike previous work that directly applies traditional signal processing methods\cite{yu2024multi} or relies on model learning capabilities\cite{wu2019concatenate}, we aim to enhance the SINR of target appliances through pre-processing by leveraging the stability and periodicity of appliance currents. We propose a feature extraction called steady-state differential (SsDiff) current, which has two key advantages.

\subsubsection{Enhanced SINR from Differential Current}

\begin{figure}[t]
  \begin{minipage}[t]{\linewidth}
  \centering
  \hspace{0.1\linewidth}
  \includegraphics[width=0.8\linewidth]{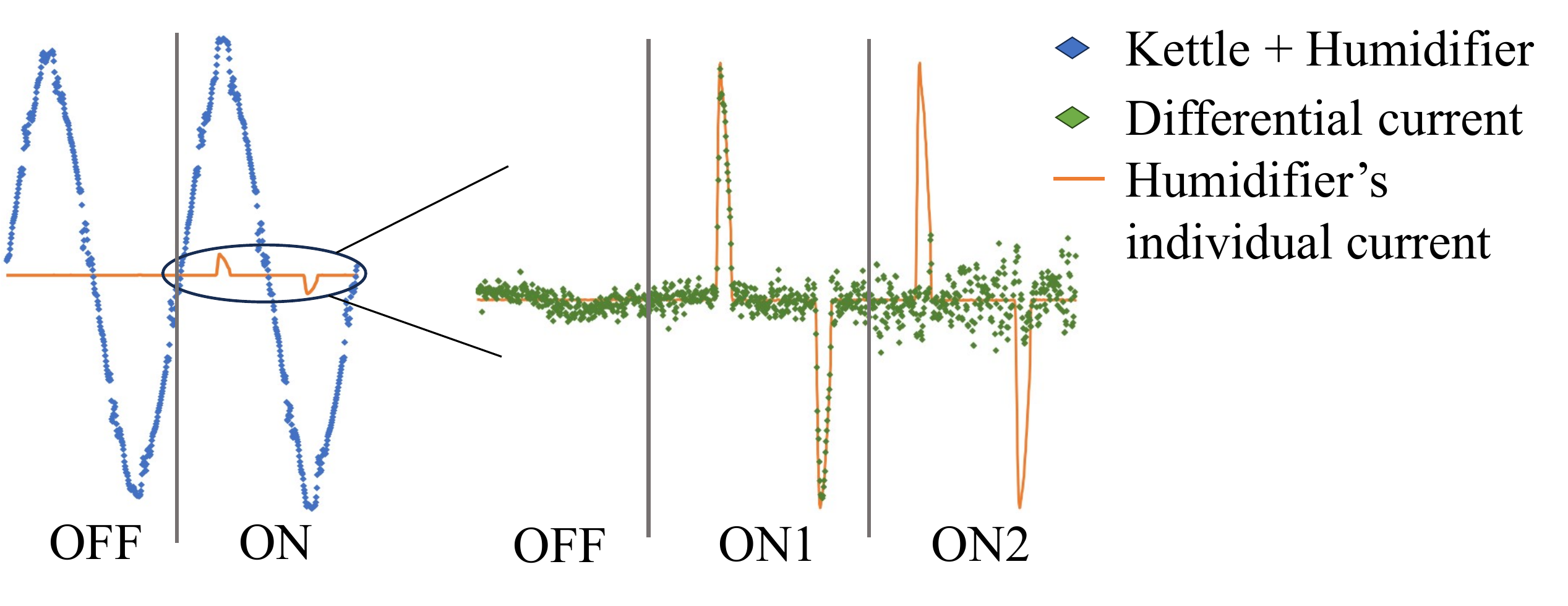}  \caption{Comparison between aggregated current and differential current. The right subplots show the humidifier's current and the differential current calculated from three differential pairs: both before the event, at the two ends of the event (ON1), and both after the event(ON2).}
  \label{fig:SsDiff-Benifit}
  \end{minipage}
  % \Description{System overview}
\end{figure}

To calculate SINR, we express the signal power as the power consumption of the appliance. Given the power of the ${i}$th appliance at time $t$ is ${P_{i,t}}$. The background noise is ${N_{B,t}}$. The SINR for the $k$th targeted appliance from raw data at time $t$ is expressed as:
\begin{equation}
    {SINR_{k,t}=\frac{P_{k,t} }{\sum_{i!=k}^{n}P_{i,t} + N_{B,t}}}
\end{equation}

Therefore, smaller power appliances are generally more difficult to recognize directly from the aggregated current when running concurrently with larger ones. However, the differential current offers a different perspective. Differential current's effectiveness relies on Kirchhoff's current law, simplified as $I_{sum}=I_{target}+I_{background}$. When background appliance currents remain relatively stable over short durations, turning on or off the target appliances should result in a proportionate change at the main meter. Given that a switch event of \textit{k}th appliance occurs within the range of (t-d, t), one of $P_{k,t-d}$ and $P_{k,t}$ corresponds to the power of the ON stage, $P_{k,\textit{ON}}$, while the other corresponds to $P_{k,\textit{OFF}}$. And $P_{k,\textit{OFF}}$ is approximately zero for most appliances. Therefore, the SINR of the kth appliance event from the differential current waveform with differential pair (t-d, t) can be expressed as follows:
\begin{equation}
\begin{aligned}
    \text{SINR}_{k,t} &= \frac{|P_{k,t} - P_{k,t-d}|}{\sum_{i \ne k}^{n} N_{i,t}' + N_{B,t}'} = \frac{P_{k,\textit{ON}} - P_{k,\textit{OFF}}}{\sum_{i \ne k}^{n} N_{i,t}' + N_{B,t}'} \\
    &\approx \frac{P_{k,\textit{ON}}}{\sum_{i \ne k}^{n} N_{i,t}' + N_{B,t}'} \label{con:sinr-definition}
\end{aligned}
\end{equation}

${N_{i,t}^{'}}$ represents the power of differential noise from the $i$th appliance. Since background appliances always remain relatively stable for a short period, ${N_{i,t}^{'}}$ is significantly smaller compared to the original power. For example, assume an indoor environment with a 40±1W humidifier and a 1500±20W electric kettle working together, with a background noise of 10±10W. The humidifier's maximum SINR is 41/1480 if we identify directly from the total current. However, after applying differential current, the minimum SINR for the humidifier's switching event is 39/(40+20) = 39/60.

To demonstrate differential current's benefits more vividly, we present an example in \autoref{fig:SsDiff-Benifit}. The left subplot shows the total current for the humidifier's on and off state when the background appliance is a kettle. Due to the high-power background kettle, the total current waveform indistinguishable changes before and after the humidifier switch. On the other hand, when background appliances are relatively stable (often the case), and the differential current pairs are distributed around the switching events, the differential current will capture the waveform of the individual current change from the aggregated current. As seen in \autoref{fig:SsDiff-Benifit}, the waveform corresponding to ON1 matches the current values of the target appliance and makes it more distinguishable. Conversely, only electrical noise remains if the differential pair is in the same appliance state combination (such as OFF and ON2 in \autoref{fig:SsDiff-Benifit}).

\begin{figure}
    \centering
    \hspace*{0.7cm}
    \includegraphics[width=0.9\linewidth]{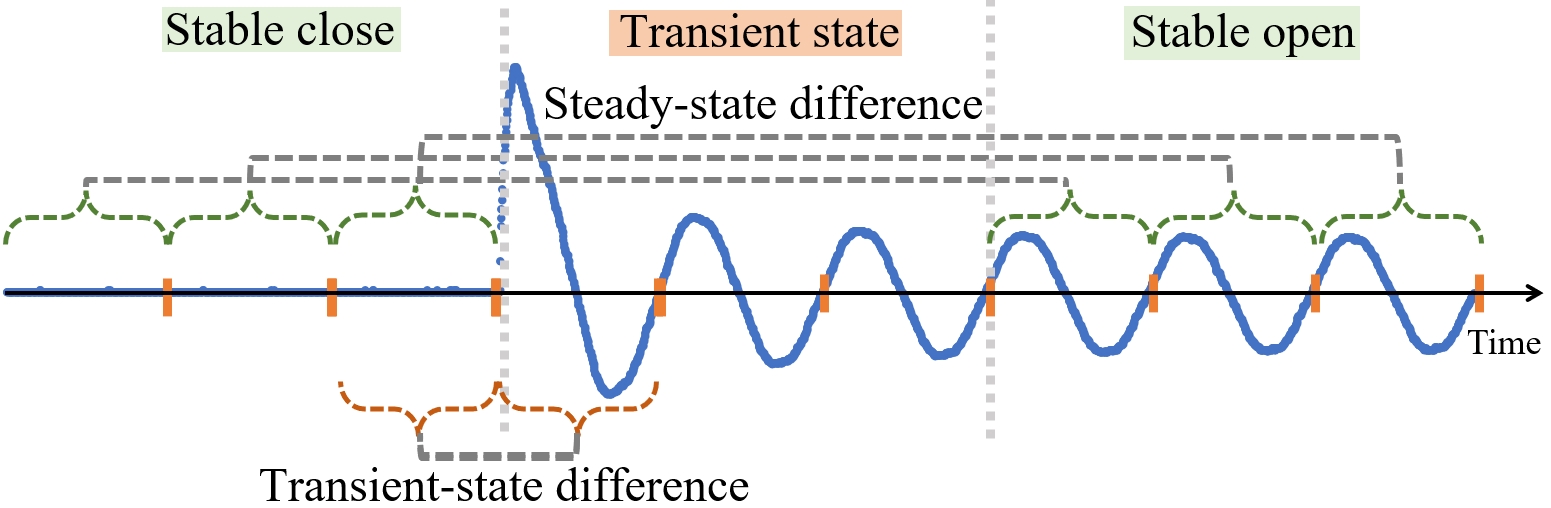}
    \caption{
    The incandescent lamp's individual current changes around turning on. When background currents are stable, the steady-state differential operation on total current is supposed to get changes in the incandescent lamp's current.
    }
    \label{fig:current-at-incandescent-lamp}
\end{figure}

\subsubsection{Enhanced Robustness from Windowed Voting.} As shown in \autoref{fig:SsDiff-Benifit}, differential total current can extract the individual current of switching appliances while background appliances are running steadily. Moreover, most appliances undergo a transient state between stable on and off states\cite{picon2016cooll}, as shown in \autoref{fig:current-at-incandescent-lamp}. Such transient current waveforms are more random and easily confused with waveforms from other background appliances. Therefore, we insert a fixed interval between differential current pairs, allowing us to bypass the transient state and directly capture the differential current between stable on and off states. 

Unlike previous work that expects models to learn differential information from raw inputs\cite{wu2019concatenate} or extract steady-state differential currents for classification after detecting events\cite{ding2020non}, we employ an one-step sliding window approach for streaming steady-state differentials, as shown in \autoref{fig:current-at-incandescent-lamp}. 
Additionally, we leverage another observed characteristic of the differential currents to enhance our algorithm's robustness further.
Given a differential interval of $D$ and an appliance transient length of $t$, the steady-state differential current can ideally obtain $(D - t)$ cycles of differential current signature corresponding to the event (as illustrated by the three pairs of differential currents in \autoref{fig:current-at-incandescent-lamp}). In other words, the classifier will output the event $(D - t)$ times. Thus, we propose a voting-based post-processing method that reports events when the report number of the classifier greater or equal to a threshold $T$. Such a process can tolerate $(D-t-T)$ false negatives and $T-1$ false positives caused by background interference, which will effectively enhance the robustness of the model predictions.

\subsection{Considerations of Model Design} \label{preliminary-of-ap}

\textbf{Problem Definition.} Different from appliance state recognition, we define appliance event recognition in as a time-series multi-class classification problem instead of a multi-label classification problem. In SustDataED2, out of 12,252 events involving 18 appliances, only two human-related events occurring within a specific second. Out of the 799 events for the eight appliances in the BLUED dataset, also only two events occur in the same 1-second interval. The statistics indicate a low collision probability for human-related events and set reference for maximum differential gap length.

\textbf{Determination of crucial parameters.} Differential interval and voting threshold are critical parameters in the algorithm pipeline. The differential interval for calculating SsDiff current should be long enough to jump over transient features but reasonable to avoid simultaneous events and rising risk of background noise.  Moreover, specific thresholds for each appliances directly affect recognition accuracy, which is determined during model training by maximizing the F1-score of event predictions on the training set.

\textbf{Classifier Selection.} We conducted preliminary experiments with various deep neural networks for time series classification\cite{ismail2019deep} and found that CNN and CNN-LSTM achieve higher accuracy and better computational efficiency. Different appliances show varying power ranges and harmonic content in their current waveforms. Hence, we combined FFT with XGBoost\cite{chen2016xgboost}, which empirically performs better on unsmooth target functions\cite{grinsztajn2022tree}. We ultimately selected these three classifiers for comparison in our evaluation.

\begin{figure}
  \begin{minipage}[t]{\linewidth}
    \centering
    \includegraphics[width=\linewidth]{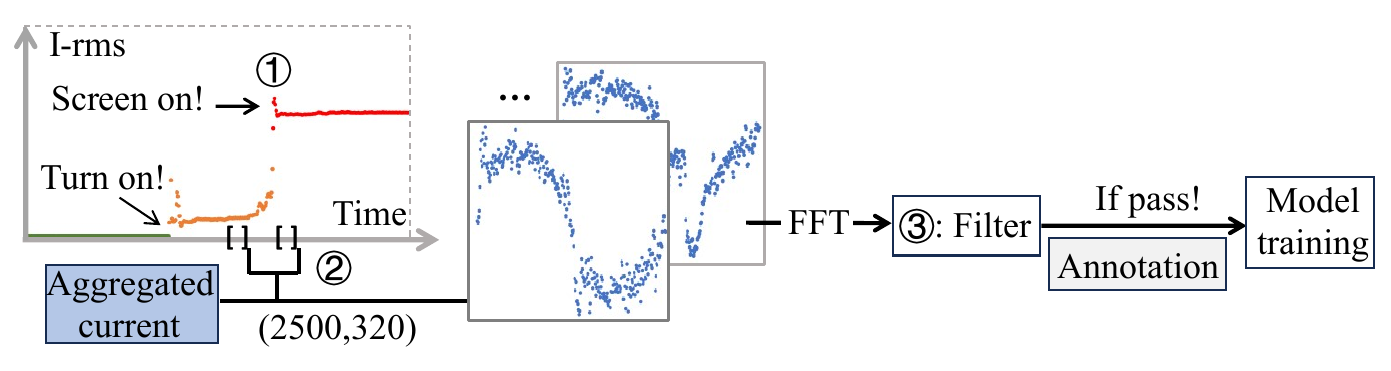}  
    \caption{Training data preparation pipeline. The smart screen is a multi-component appliance and undergoes two events after switch on: \textit{system boot} and \textit{screen on}, the second one is more noticeable since the screen costs more energy.}
    \label{fig:pre-train-pipeline}
  \end{minipage}
\end{figure}

\subsection{Algorithm Pipelines}
The differential currents of events of low-power appliances are vulnerable to fluctuations caused by background appliances, so the pipeline design for model training and inference must account for noise interference with recognition.

% \textbf{Obvious Event Point Localization.} Some appliance events consistently go through several stages. For example, a smart screen's power-on sequence includes two distinct phases: system boot and screen activation. The screen consumes more power, resulting in a more noticeable change in current before and after the screen is turned on. Unlike strictly accurate dataset annotation,  we choose more noticeable current change points as their on/off events to predict(such as S-Screen's \textit{screen on} event). Therefore, we raise the RMS-current thresholds to automatically locate more apparent event positions when we prepare training data.

\subsubsection{Training Data Preparation} \label{Data-prepare}
To reduce the effect of contaminated data on training and achieve data augmentation. We propose a three-step pipeline (depicted in \autoref{fig:pre-train-pipeline}) to automatically extract and filter a sufficient number of SsDiff currents with corresponding labels for model training.

\begin{enumerate}
\item \textbf{Obvious Event Localization}: Unlike strictly accurate dataset annotation, we choose more noticeable current change points during inevitable stages as their on/off events to predict(such as Smart Screen's \textit{screen on} event in \autoref{fig:pre-train-pipeline}). Therefore, we raise the RMS-current thresholds to automatically locate more apparent event positions to prepare training data.
% Locate more noticeable current change points during inevitable stages after appliance on/off events (such as S-Screen's \textit{screen on} event in \autoref{fig:pre-train-pipeline}). Therefore, we raise the RMS-current thresholds to locate more apparent event positions automatically than the actual OFF-ON positions from data annotation.
\item \textbf{Data Augmentation}: Steady-state differential currents are obtained by performing pairwise differential operations on the current from 50 stable ON and 50 stable OFF states on either side of appliance switches, generating 2500 pairs of differential current differences for each appliance event.
\item \textbf{Sample Filtering}: The enhanced samples are filtered by comparing the fundamental frequency harmonics with the individual appliances', reducing contaminated samples.
\end{enumerate}

% In the end, the filtered training data is used to train the classifiers, which are integrated in algorithm pipeline in next subsection.

\subsubsection{Inference pipeline}
\begin{figure}[t]
    \begin{minipage}[t]{\linewidth}
        \centering
        \includegraphics[width=\linewidth]{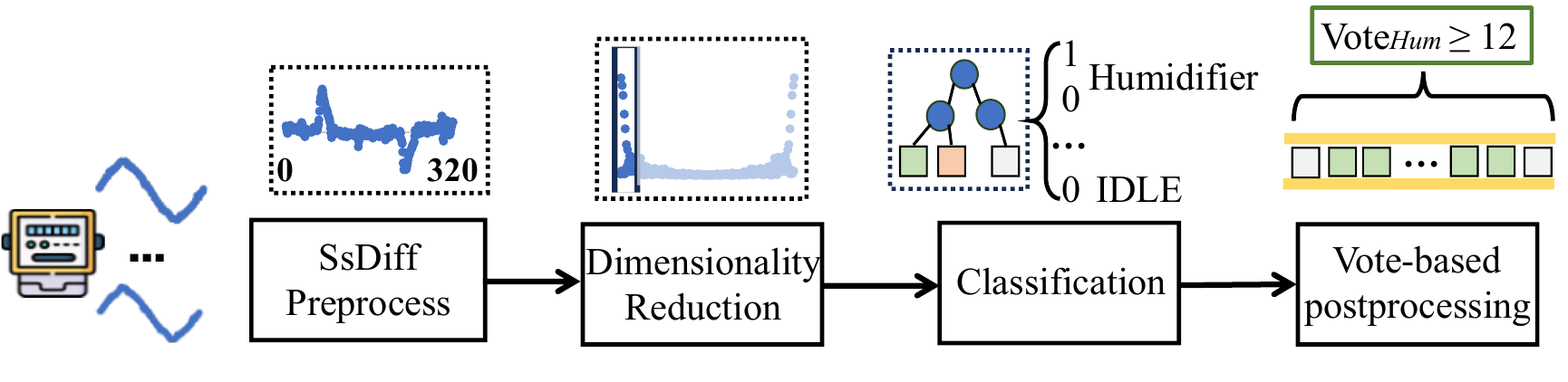}
        \caption{The inference pipeline of the event recognition algorithm. The pipeline's input is read from main meter. When vote number is over threshold, the event is reported.}
        \label{fig:event-recognition-pipeline}
    \end{minipage}
\end{figure}

Hawk's inference pipeline follows a basic preprocess-classifier-postprocess structure. During inference, Hawk's algorithm pipeline reads the aggregated current, organized by voltage cycles, from the smart meter to detect and classify appliance events. Our algorithm pipeline based on FFT-XGBoost is shown in \autoref{fig:event-recognition-pipeline}, while the pipelines based on CNN and CNN-LSTM use the raw current waveform as input.

\begin{enumerate}
    \item \textbf{Differential Current Calculation}: We compute Steady-state Differential currents by direct cycle-level current subtraction from a 30-cycle circular buffer.
    \item \textbf{Dimensional Reduction}: We take the first ten harmonics, including the real part, imaginary part, and magnitude, of differential current as model inputs after FFT transformation.
    \item \textbf{Classification}: Reduced features are fed into a big XGBoost model to identify events. The model outputs 37 states, including OFF-ON states of 18 appliances and an IDLE state.
    \item \textbf{Post-processing}: A post-processing strategy involves voting within an 30-cycle window. The most likely event, surpassing the voting threshold (e.g., 12 for the humidifier turn on event), is reported eventually.
\end{enumerate}

We can use the \textit{second step} to reduce the dimensionality of features while preserving sufficient information of the raw waveform because most of the current waveform energy of appliances is more concentrated in the low-frequency region. Such an operation can also eliminate the effect of high-frequency noise. Because Hawk performs continuous inference, the \textit{classifier} includes an IDLE state, indicating no events detected.
The results of the streaming classification are buffered in a queue of length the same as the differential gap. Each time, we count the occurrences of each category in the queue, and when the count of a specific category exceeds its corresponding threshold, the classifier outputs that category. This postprocessing efficiently reduces false predictions caused by noise.

\section{Prototype And Validation Settings} \label{UseCase}
In this section, we first present the design of our sampling hardware prototype and then introduce different settings: laboratory, in-the-wild study in the residual and office area with a brief description of the collected dataset.
% Parts of HawkDATA are publicly accessible at \url{https://www.kaggle.com/datasets/zijianwang01/hawkdata}, and will be published later.

\subsection{Hardware Prototype}
\begin{figure}
  \begin{minipage}[t]{\linewidth}
    \centering
    \includegraphics[width=0.8\linewidth]{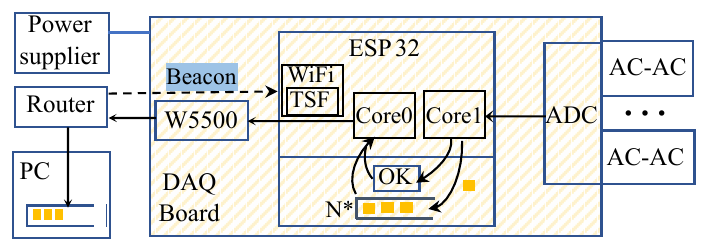}
    \caption{Hardware architecture of Hawk data acquisition. Voltage and current are collected by data acquisition boards through AC-AC sensors and transmitted via Ethernet to a data logging PC.}
    \label{fig:DAM-system-overview}
  \end{minipage}%
\end{figure}

\begin{figure}
  \begin{minipage}[t]{\linewidth}
    \centering
    \includegraphics[width=0.7\linewidth]{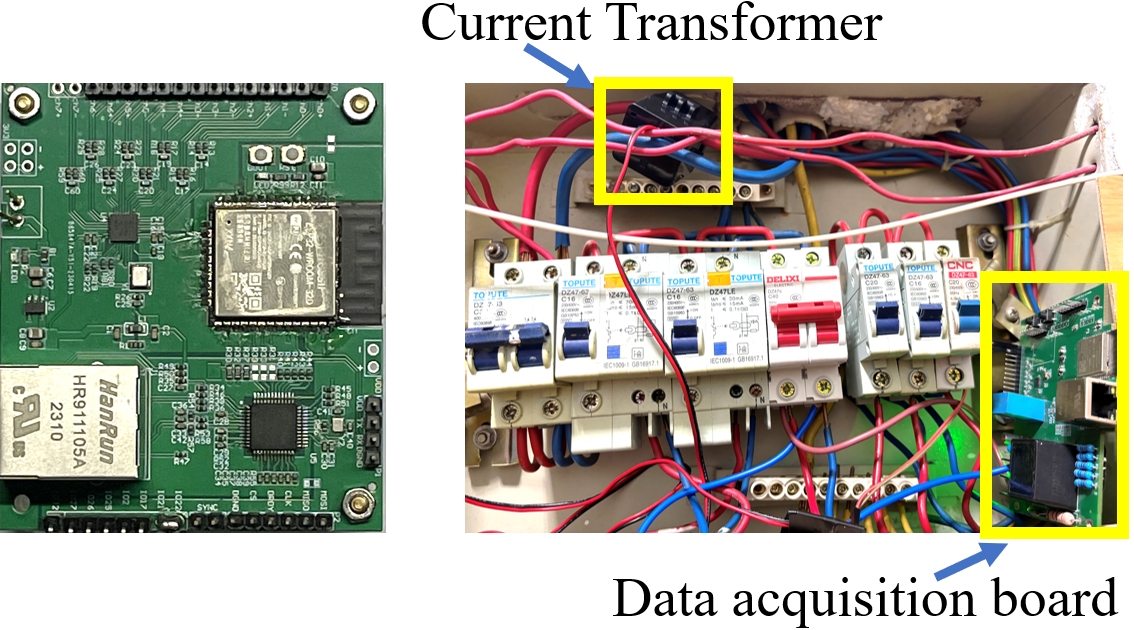}
    \caption{Two version of hardware prototypes. The left subplot shows the laboratory version powered by an external shared power supply. The deployed version in real environments will extend an AC-DC power module to avoid the need for battery maintenance.}
    \label{fig:Hawk-as-socket}
  \end{minipage}%
\end{figure}

Hawk hardware architecture, depicted in \autoref{fig:DAM-system-overview}, integrates three primary elements: the data acquisition board for high-frequency data capture, AC-(to)-AC sensors for current and voltage sensing, and a data recording infrastructure. This fundamental hardware architecture design is shared between the development and deployment stages. We developed two versions of the prototype corresponding to two stages, as shown in \autoref{fig:Hawk-as-socket}.

The data acquisition board consists of three components: a 24-bit ADS131M08 ADC for 16KHz simultaneous 8-channel sampling, an ESP32 MCU offering dual-core processing with integrated WiFi for TSF counter queries, and a W5500 Ethernet controller to transmit data reliably during dataset construction stage. 
To ensure stable and continuous sampling at 16KHz, as depicted in \autoref{fig:DAM-system-overview}, we assign the task of responding to ADC interrupts to Core1, which is free of default interrupts. Core1 reads and packages data by voltage cycles (the yellow squares in \autoref{fig:DAM-system-overview}). Conversely, Core0 is responsible for querying packet status and transmitting the cycle's data from the shared memory to the logging PC.

The AC-AC sensor unit comprises voltage and current sensors. Current sampling has three typical types of sensors\cite{ziegler2009current}: shunt resistors, Hall effect sensors, and current transformers. 
We opted for contactless current transformers for their circuit non-intrusiveness, ease of deployment, and electrical safety. 
% However, typical current transformers (CT) always have a cutoff frequency\cite{ziegler2009current}, and CT with a higher cutoff frequency costs more. We perform preliminary tests on appliances to determine the frequency domain range of the currents. Appliances with higher frequency current are measured by CTs with higher cutoff frequency, while others are measured by more cost-effective CTs. Multiple AC-AC sensors share a single data acquisition board to reduce the average sampling cost.

The collected data are streamed through a wireless router. The wireless router facilitates the Ethernet connection between the data logging PC and the sampling boards via the MQTT protocol. It also synchronizes the TSF in each MCU for the sampling synchronization strategy. We utilize a desktop as the data logger in the laboratory environment. In real-household conditions, we employ an embedded computer as the collector to reduce disruption to the household's current and reduce costs. Additionally, the collected data is stored on an external hard drive.

The current transformer selected for the prototype at the main meter is rated at 4000:1, with a shunt resistor value of 10 ohms. The ADC reference voltage is 1.25V, featuring a resolution of 24 bits and a sampling frequency of 16 kHz. Ignoring sampling deviation, the theoretical resolution of the sampling is 59.6 µA. However, the sampling accuracy is constrained by the effective number of bits of high-frequency sampling and is also affected by the nonlinear conversion of the current transformer. Therefore, in the evaluation section, we assess the linearity of the current conversion.

The hardware design is designed to maximize component utilization, with multiple AC-AC sensors sharing a single data acquisition board and taking full advantage of dual-core performance. As a result, the average sampling cost per current channel based on the data acquisition board is reduced to \$4.12, almost one-third of the cost of the Hz-level sampling board\cite{debruin2015powerblade}. This is remarkable, as the high cost of large-scale training data collection is a key obstacle for supervised method of NALM\cite{zhu2023combining}. Even in the deployment phase, this optimized design can be taken in environments with multiple smart meters, such as green communities.
% As a result, the average sampling cost of the laboratory version (as shown on the left in \autoref{fig:Hawk-as-socket}) for development stage is reduced to \$4.12, almost 1/3th of previous low-frequency prototype

\subsection{Laboratory Setting and HawkDATA} \label{Hawk-dataset}
We build a data collection laboratory to deploy 22 common appliances for data collection. Among these, 18 are programmable for future algorithms to recognize, as listed in \autoref{tab:Count-of-event-number}. The other 4 appliances, such as a wireless router and refrigerator, are constantly on, free of human activity, and used as background appliances. We select diverse appliance representations, including those with similar functions but with different electronic circuits (e.g., LED, incandescent, and fluorescent bulbs for lighting). The electrical characteristics of these appliances span a wide range, from a 5W camera to a 2160W air heater and coverage of most front-end circuits\cite{he2012front}. Besides diversity, recognition challenges are also considered, and we also choose appliances with the same working principle but varying power ratings and appliances with close power rates. This is also an advantage of HawkDATA over others from real households with biased collections of appliances.

HawkDATA is collected in the laboratory environment, with programmable appliances executing the event schedules mentioned in \Autoref{GRBGCode}. Each appliance state combination is planned to last for 20 seconds. Ultimately, the effective collection duration of our dataset is 32.2 hours, comprising a total of 5,796,650 voltage cycles for a 50Hz AC circuit. HawkDATA includes both raw and annotated versions stored in NPZ file format. The raw data version records the total current, voltage, and individual current of 18 appliances, all sampled at 16kHz. The raw-version file size is 127.8GB. The annotated version organizes and labels the data according to voltage cycles. For a 50Hz AC circuit, each cycle contains 320 points.

\subsection{In-the-wild Settings}
\begin{figure}
    \centering
    \includegraphics[width=\linewidth]{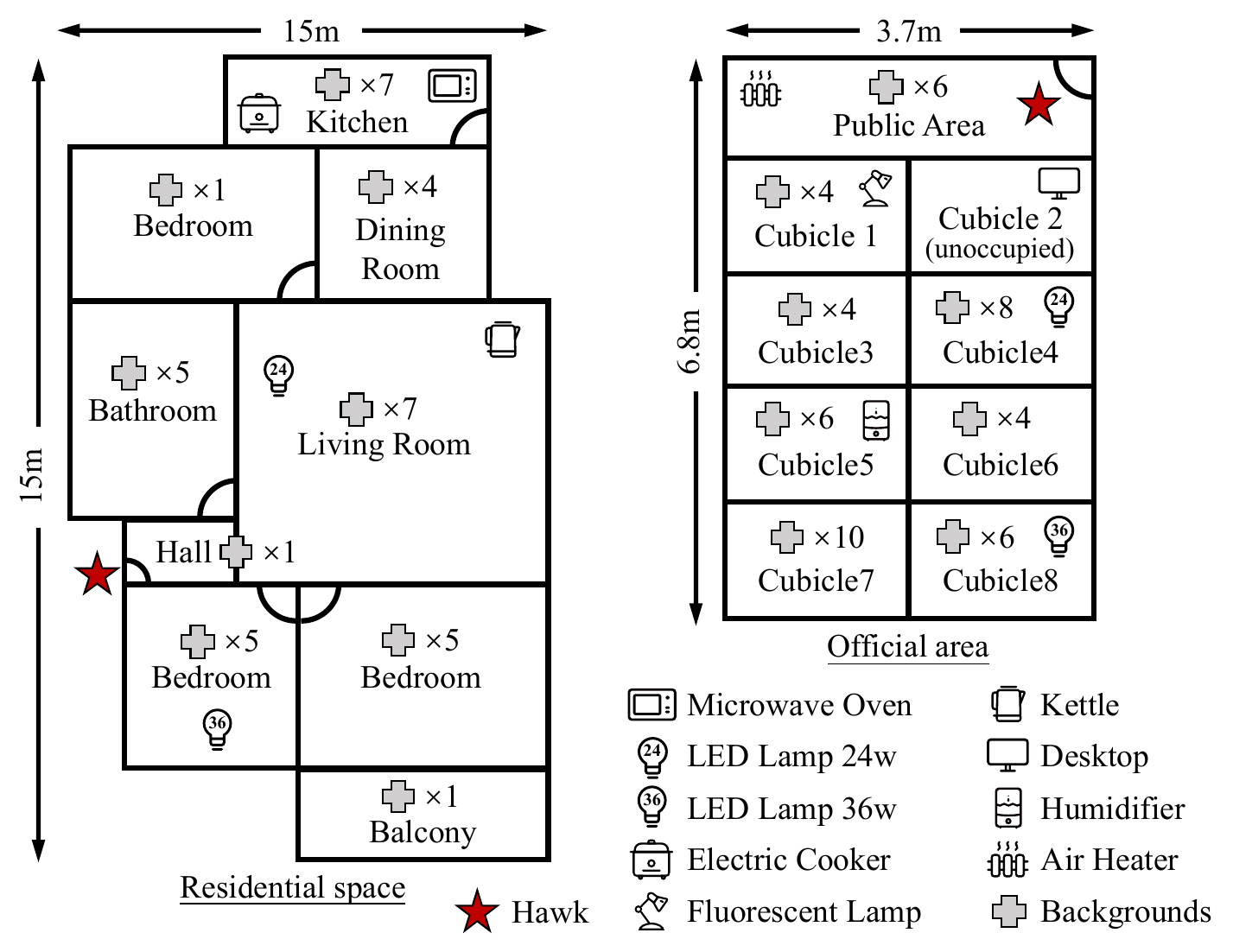}
    \caption{Deploy details of two in-the-wild settings.}
    \label{fig:In-the-wild-layout}
\end{figure}

The field experiment aims to validate the Hawk system and ensure that it aligns with the application scenarios. The Hawk system, with a single sensor for inference (shown on the left of \autoref{fig:In-the-wild-layout}), was tested in residential and office spaces, and target appliances subsets from HawkDATA were repurchased and deployed without affecting the original appliances. Including data collection laboratory, three places were selected from three different regions, with a total distance exceeding 2000 km between the points. All three regions operate under a 220V, 50Hz AC standard, but there are slight differences due to different electrical infrastructures and varying electricity consumption trends.

For the residential environment, we selected a 103-m$^2$ three-bedroom apartment with three inhabitants and 36 unknown background appliances, of which more than 40\% were lighting devices, totaling more than 42 operating states. We deployed five representative appliances from the HawkDATA set: two LED lights (24W and 36W), an electric cooker, an electric kettle, and a microwave oven, for seven days. Observations revealed the use of 35 of 36 background appliances during this period.
For the office environment, we chose a 25-m$^2$ graduate student office with seven occupants and 48 appliances, totaling more than 51 operating states. Since most office appliances are less than 150W, we selected low-power appliances from HawkDATA, including a desktop, two 24W and 36W LED lights, a humidifier, a fluorescent lamp, and a high-power air heater as a background appliance when few appliances were used. Observations showed that 44 of 48 background appliances were used during this period.

The deployment of the Hawk model is in a manner of application scenarios mentioned in the Introduction section, which are deployed without modification. The inference program is implemented in C and is attached to a physical core of embedded computers (Jetson Orin NX 16GB) to enable real-time inference. It uses the MQTT protocol to subscribe to continuous data generated by a single sensor node at the main meter, as shown in \autoref{fig:In-the-wild-layout}. Targeted appliances are manually switched on and off, with both switch actions and model recognition results recorded, including false positive samples when no switch action occurs. The switching frequency is adjusted according to the number of occupants, set to 15, 30, or 60-minute intervals. In addition, the deployed appliances do not interfere with the operation of original appliances, ensuring realistic and diverse background current.

\section{EVALUATION} \label{evaluation}

\subsection{Data Acquisition Quality Evaluation}
As a typical distributed data collection, the quality of NALM data is evaluated from two dimensions: sample synchronization accuracy between sensor nodes and single-point current conversion linearity.

\subsubsection{Sampling synchronization evaluation}

\textit{\\Baseline.} The baseline selected for comparison is the Time Synchronization Function (TSF) of the 802.11 protocol, which can achieve microsecond synchronization accuracy\cite{chen2021understanding, mahmood2014impact}. Additionally, ESP32 provides a TSF interface, making it a suitable baseline for our low-cost distributed solution.

\begin{figure}[t]
  \begin{minipage}{\linewidth}
    \centering
    \includegraphics[width=0.8\linewidth]{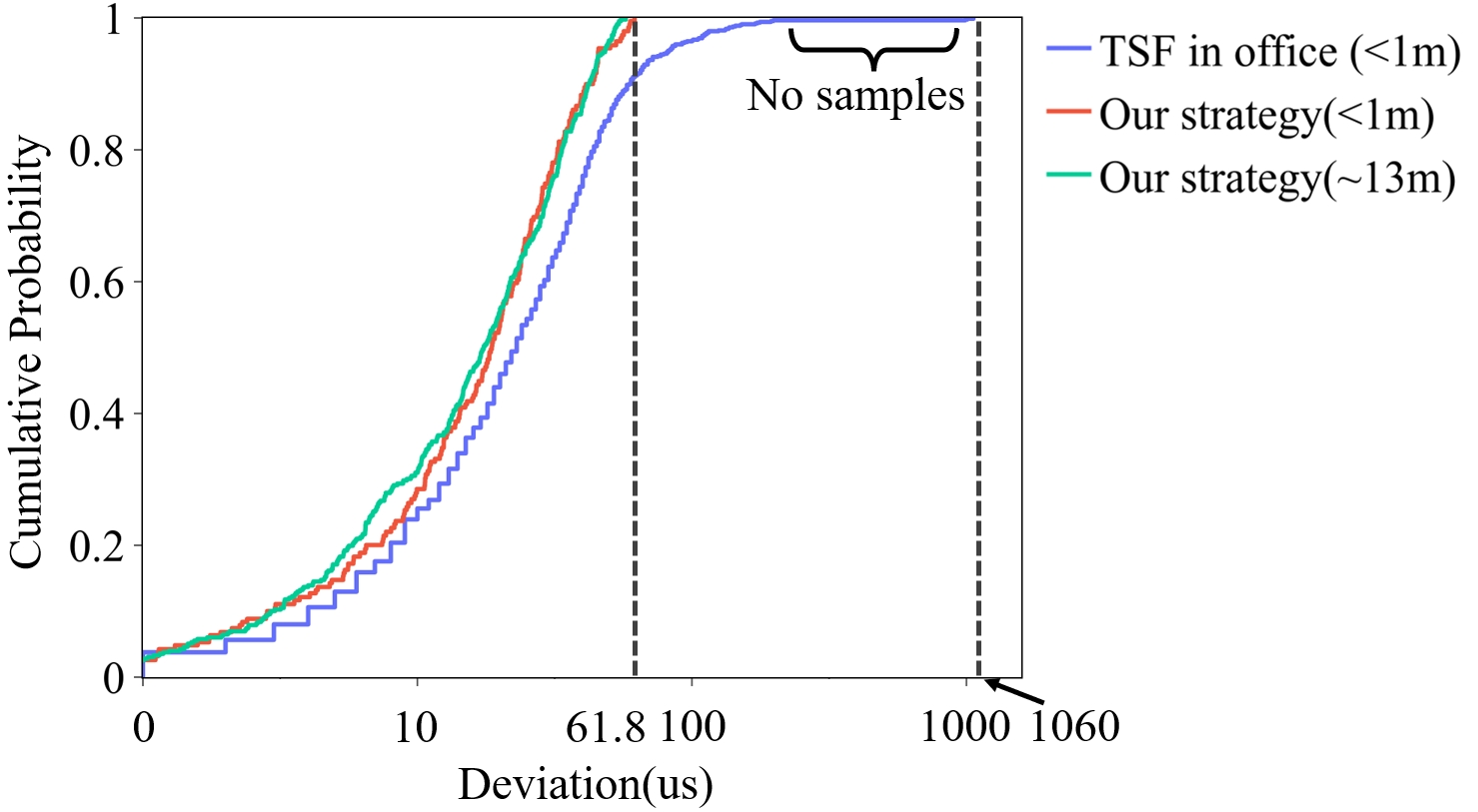}
    \captionsetup{justification=raggedright, singlelinecheck=false}
    \caption{CDF of synchronization error of our strategy and
    TSF. Our CDF is smoother due to continuous timestamp.}
    \label{fig:cdf-of-synchronization-error}
  \end{minipage}%
\end{figure}

\textit{Evaluation method.}  We assess synchronization error by comparing the timestamps of the same current surge point recorded by two nodes. Incandescent lamps experience a current surge when switched on. Such transient current surge is simultaneously detectable by the main meter and the submeter attached to the incandescent lamp. We switch on and off the incandescent bulb using a relay to collect the synchronization deviation. Two meters for testing are placed less than 1 meter from the accessed wireless router to test the best performance of TSF synchronization. Additionally, a distance of around 13m is set to evaluate the impact of electrical circuit length on SPT. The wireless router used for the experiment is an Asus AC66U-B1. The deviation of the SPT synchronization strategy is a continuous phase value, while the TSF-based synchronization strategy has a resolution of 1 microsecond.

\textit{Result Analysis.} The result in \autoref{fig:cdf-of-synchronization-error} shows that Hawk's synchronization strategy based on SPT can achieve an average synchronization accuracy of 20.60us, improved by 59.2\% compared to TSF, averaging 32.8us. Our maximum synchronization error is 61.80us, about 1/17.2 of TSF. 
We can see a gap in the error distribution of TSF; 99.65\% of the synchronization accuracy is within 220us, while the other 0.35\% are distributed from 980us to 1060us. This gap is due to the 1-ms WiFi packet jitter\cite{thi2022ieee}. 
Our experiments show that the jitter ratio of beacon packets is affected by wireless channel interference and the distance between the AP and the STA.
In our testing scenario, several wireless APs in the office area are the sources of interference. On the other hand, our proposed SPT strategy is more robust. When the two sensor nodes are 13 meters apart, the overall accuracy improves. This is due to introduced random quantification errors, by equating the switch events of a continuous position of the events to a sampling point. The more comprehensive analysis of our synchronization error will be presented in future work.

\subsubsection{Current conversion linearity evaluation}
\begin{figure}[t]
  \begin{minipage}{\linewidth}
    \centering
    \includegraphics[width=0.8\linewidth]{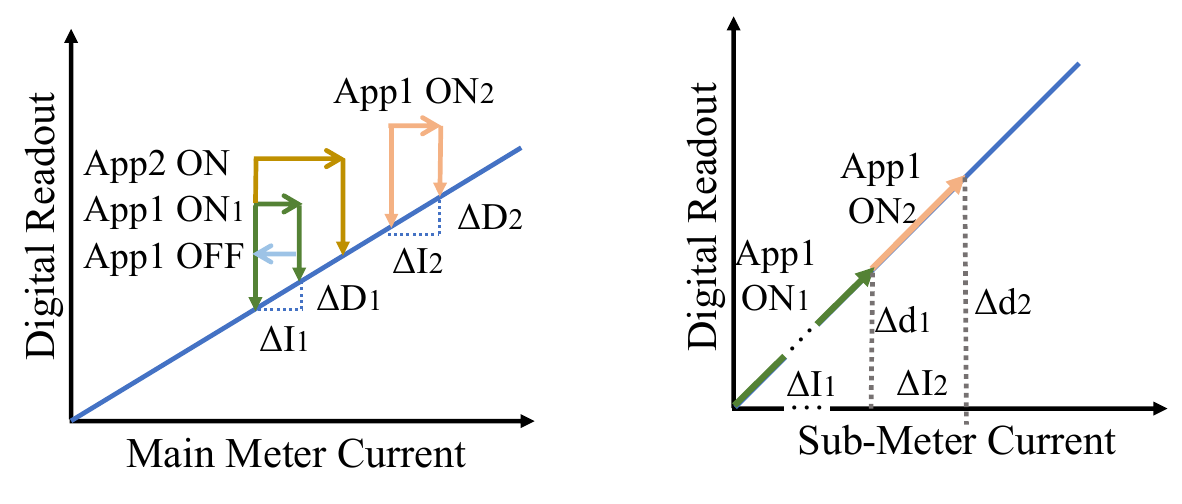}
    \caption{Current conversion differential linearity of different events with different background current. The ratios are calculated from diverse situations.}
    \label{fig:linearity-calculation}
  \end{minipage}%
\end{figure}

\begin{figure}[t]
  \begin{minipage}{\linewidth}
    \centering
    \includegraphics[width=0.8\linewidth]{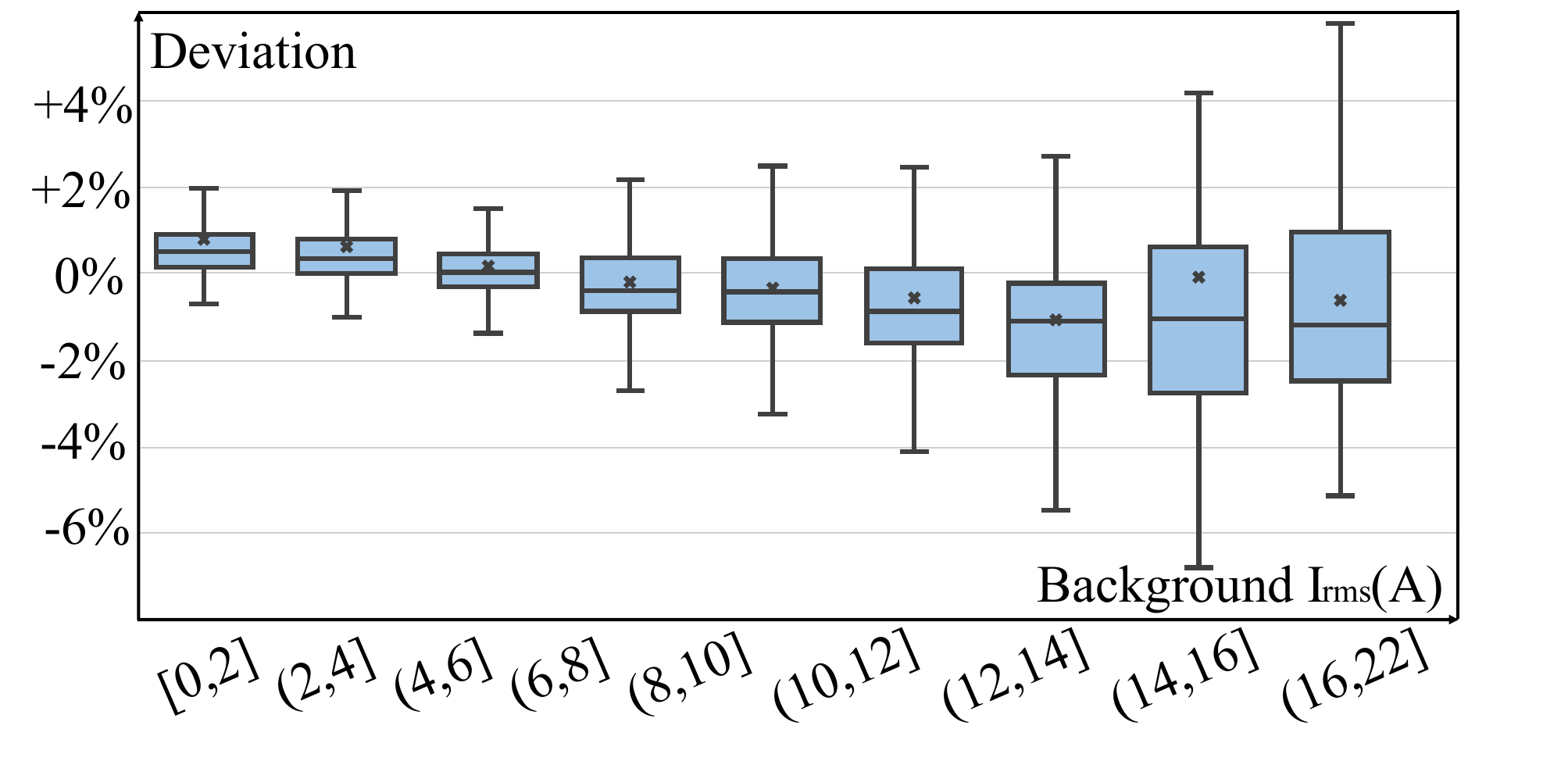}
    \caption{Differential linearity of current conversion in HawkDATA.}
    \label{fig:linearity-result}
  \end{minipage}%
\end{figure}

\textit{\\Evaluation method.}
One of the challenges for sampling accuracy evaluation experiments is that we lack an ideal measurement for high-frequency current sampling instruments as a standard. Such standards invariably have their deviation ranges and come with prohibitive costs. Inspired by a commonly used evaluation metric of ADC/DAC, differential linearity, we designed an assessment method. Differential linearity of ADC refers to the stable relationship between the digital output changes and the analog input variations at different starting points. As shown in \autoref{fig:linearity-calculation}, we evaluated the differential linearity of HawkDATA by comparing the stability of the ratio between the total meter reading differences and current differences before and after switching events for different appliances under varying background currents. The current difference is presented as sub-meter reading differences. Since most appliances operate within a more stable and lower current range when compared with aggregated current, we assume that the current differences maintain a consistent ratio with the reading differences. Finally, we organize different appliance switches’ average normalized linearity rates within the same range of background currents into a box and show their changes along the change of background currents in \autoref{fig:linearity-result}.

Ensuring the differential linearity of our sampled values also enables our proposed steady-state differential current-based feature extraction. Though not guaranteeing that the acquired data represents exact values, differential linearity ensures a stable proportional relationship. Such deviations are often deemed inconsequential after typical pre-processing steps like normalization.

\textit{Result.} According to \autoref{fig:linearity-result}, both mean and median values deviate by no more than 2\% from the ideal unity, substantiating the stability of our data conversion process. Since this metric is statistics of raw HawkDATA, the final results ensure that within the HawkDATA, the average reading deviation of the same current variation, influenced by background current, does not exceed 2\%. The data visualized in \autoref{fig:linearity-result} reveals two phenomena with increased background current. The first observation is a slight reduction in the average normalized linearity rate which may related to property of current transformer. The second observation is the widening spread of the linearity rate distribution as the background current intensifies. The broadening is likely a consequence of the increased electrical noise and interference associated with a more significant number of active background appliances, affecting the calculation of differential linearity rates.

\subsection{Dataset Construction Evaluation}
Although the sequence generated from group-randomized balanced Gray code guarantees balance, the practical execution of programmable appliances may deviate from this due to factors like the timed switching or operational failures; for instance, the microwave oven in \autoref{tab:Count-of-event-number} exhibits more frequent switching.

\textit{Baseline.} Our baseline is SustDataED2\cite{pereira2022sustdataed2}, a high-frequency NALM dataset collected from a real household. SustDataED2 comprises the same number of appliances as the programmable appliances in Hawk, making it a reasonable baseline as representative of the traditional NALM dataset construction schema.

\begin{table}
  \caption{The number of events and on-state cycles for the appliance in the HawkDATA.}
  \label{tab:Count-of-event-number}
  \tabcolsep 3pt 
  \scalebox{0.8}{
  \begin{tabular}{ccc|ccc}
    \hline
    \textbf{Appliances}&\textbf{Event}&\textbf{On-state}&\textbf{Appliances}& \textbf{Event}&\textbf{On-state}\\
    \hline
    Monitor&302&1,774,054&SmartScreen&294&1,780,251\\
    Humidifier&278&1,694,105&FluorescentLamp&292&1,798,803\\
    LEDLamp24w&278&1,666,026&Television&290&1,707,688\\
    IncandescentBulb&296&1,856,095&Washer&288&1,693,044\\
    ElectricCooker&296&1,656,548&MicrowaveOven&1310&1,628,551\\
    InductionCooker&288&1,705,172&AirHeater&282&1,712,366\\
    LEDLamp36w&280&1,718,433&ElectricKettle&286&1,687,541\\
    Stirrer&290&1,651,833&PhoneCharger&294&1,793,826\\
    Desktop&286&1,827,268&SweepingRobot&292&1,695,928\\
    \hline
  \end{tabular}
        }
\end{table}

\textit{Metrics.}
The metric to measure dataset balance is balance ratio (BR), the inverse of a commonly used imbalance ratio(IR)\cite{Ye2021ICCV}, to avoid the minimal value of zero\cite{filip2011blued}. The category BR is defined as,
\begin{equation}
    BR = \frac{N_{min}}{N_{max}}
\end{equation}
where $N_{min}$ is the sample size of the minoriest category and $N_{max}$ is the sample size of the majoriest category. What's more, there are different balance requirements for state and event recognition algorithms. The balance requirement of the event recognition algorithm entails a uniform distribution of switch events across different appliances, maximizing the category BR. In contrast, state recognition algorithms require balance in two aspects\cite{tarekegn2021review}. First, the number of ON and OFF state samples for each type of appliance should be as close as possible. Second, the number of ON states between different appliances should be balanced.

As for the metrics of dataset diversity, we use the number of state combinations to represent the diversity of states and qualitatively represent event diversity by a heatmap of the appliance on/off distributions with different levels of background RMS current.

\textit{Result Analysis.} \autoref{tab:Count-of-event-number} records the distribution of our event and state combinations. Most appliances have a consistent number of collected events and ON states, except that the microwave oven's timed shutdown feature caused many unplanned switch events during data collection. Ultimately, HawkDATA's category BR of the event is 0.212, significantly higher than that of SustDataED2, which is 0.0014. And the category BR of ON states across appliances is 0.877, exceeding SustDataED2's 1.22*$10^{-4}$.
We use the average ON-OFF BR as the evaluation metric of the overall balance between ON and OFF states. HawkDATA's average ON-OFF BR is 0.424, surpassing SustDataED2's 0.338.

\begin{figure}[t]
  \begin{minipage}{\linewidth}
    \centering
    \includegraphics[width=\linewidth]{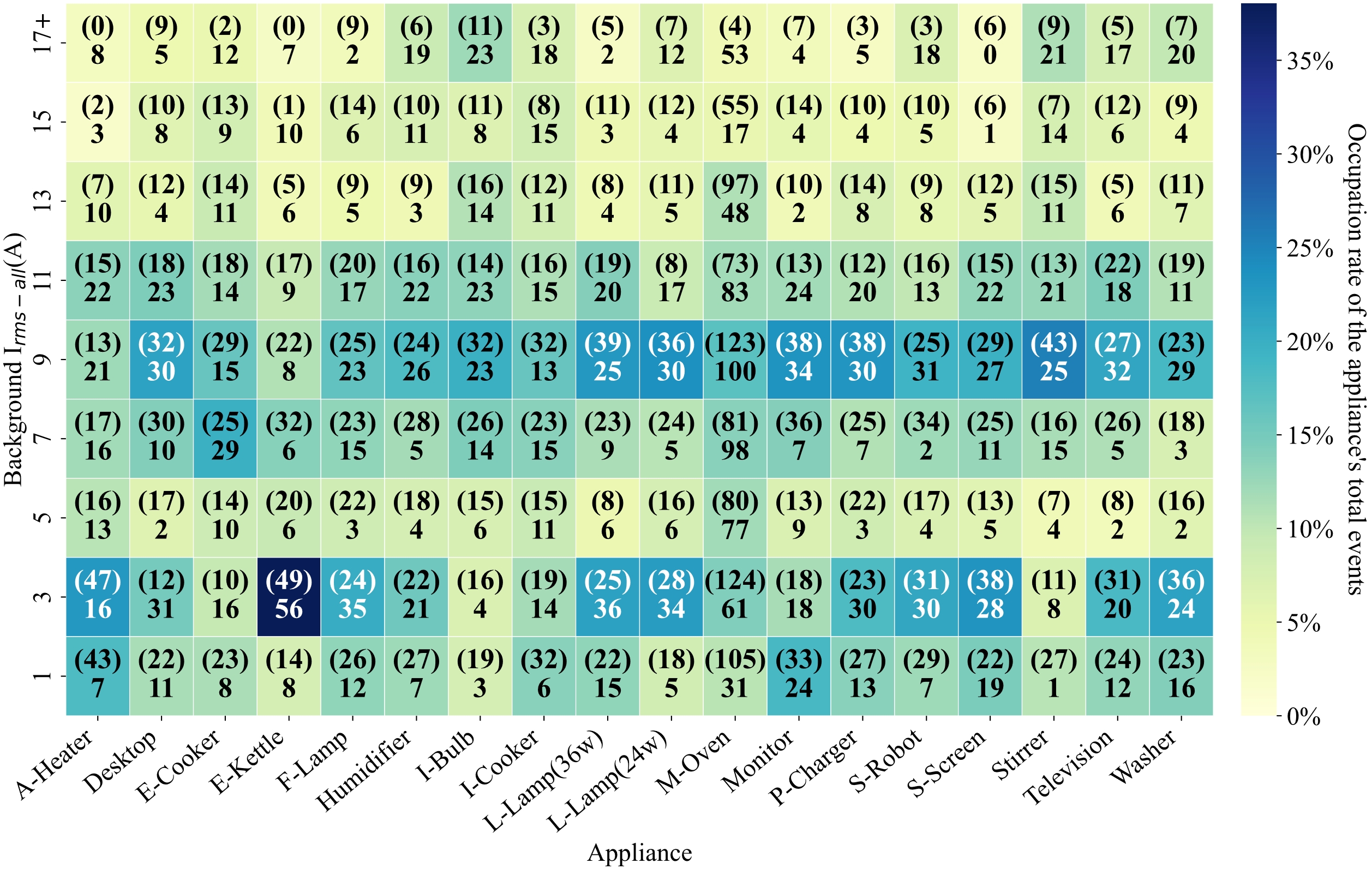}  
    \caption{Each cell contains two numbers: the top number denotes the count of events occurring in the training set, and the bottom number corresponds to the testing set.}
    \label{fig:Event-diversity-evaluation}
  \end{minipage}
\end{figure}

As for event diversity evaluation, all appliances exhibit switch events across all current ranges in \autoref{fig:Event-diversity-evaluation}. As for state diversity, HawkDATA records 4,558 unique state combinations across data collection of 32.2h. As a comparison, SustDataED2, a dataset collected from real households, acquired only about 718 unique states in 2,304 hours. HawkDATA records a unique state every 25.43 seconds and a switch event every 18.91 seconds during data collection. The number of unique state combinations collected per unit of time, which we refer to as the diversity density or information density of the dataset, is much higher than others, reflecting the efficiency of Hawk dataset construction.

\subsection{Algorithm Pipeline Evaluation}

\subsubsection{Overall Setup} The evaluation of our algorithm pipeline consists of two parts: first, a set of ablation experiments targeting the three key designs of the Hawk system; second, a comparison of event classification accuracy on the BLUED dataset, which includes multiple multi-state appliances under different AC settings.

We design three ablation experiments focusing on the data balance, classifier performance, and steady-state differential current of the Hawk system. The first experiment compares event recognition accuracy across different classifiers trained on balanced and imbalanced HawkDATA. The second experiment adjusts the differential interval to evaluate its impact on recognition accuracy. The last experiment evaluates the effect of differential current processing on state recognition accuracy and compares it with open-source state recognition algorithms\cite{he2023msdc, wang2021calm}.

\textit{Baselines}.
\textit{MSDC}\cite{he2023msdc} uses CNN to recognize appliance OFF-ON states from low-frequency power signals, followed by state corrections with conditional random fields (CRF). 
\textit{CALM}\cite{wang2021calm} is based on high-frequency signals and takes BiLSTM to infer appliance states from the raw high-frequency current.
\textit{MTS-Shapelet}\cite{yu2024multi} employs a multi-time-scale shapelet to process high-frequency current waveforms. It provides detailed accuracy results for appliance classification events, facilitating the comparison of different appliances.

\textit{Datasets}.
\textit{HawkDATA}, as previously discussed in \Autoref{Hawk-dataset}, is collected in a laboratory environment with 220V, 50Hz AC power. The average power is calculated from the total current and voltage to generate a labeled dataset for MSDC based on low-frequency signals.
The \textit{imbalanced HawkDATA} is resampled based on the relative proportions of appliance usage in actual households (\cite{filip2011blued,pereira2022sustdataed2}, as shown in the left part of \autoref{fig:compare-classifier-and-balance-effect}), keeping the total event count constant. We add differential current during periods without appliance events to increase the sample size for underrepresented types, based on Kirchhoff's current law.
\textit{BLUED}\cite{filip2011blued} is a high-frequency (12KHz) NALM dataset collected in a residential environment with 240/120V, 60Hz AC power over seven days. Some of collected appliances (such as fridge) are multi-state appliances.

\textit{Metrics}. We use the F1 score to evaluate recognition accuracy and average F1 score to evaluate overall performance on balanced HawkDATA. For the imbalanced BLUED dataset, we also employed the weighted average F1 score\cite{yu2024multi}. However, we still prefer the average F1 score. This is because the weighted F1 score disproportionately emphasizes appliances like fridge, which have frequent automatic switches but are less related to human behavior.

\subsubsection{Impact of classifier and dataset balance} \label{Event-evaluation}
\begin{figure}[t]
  \begin{minipage}{\linewidth}
    \centering
    \includegraphics[width=0.9\linewidth]{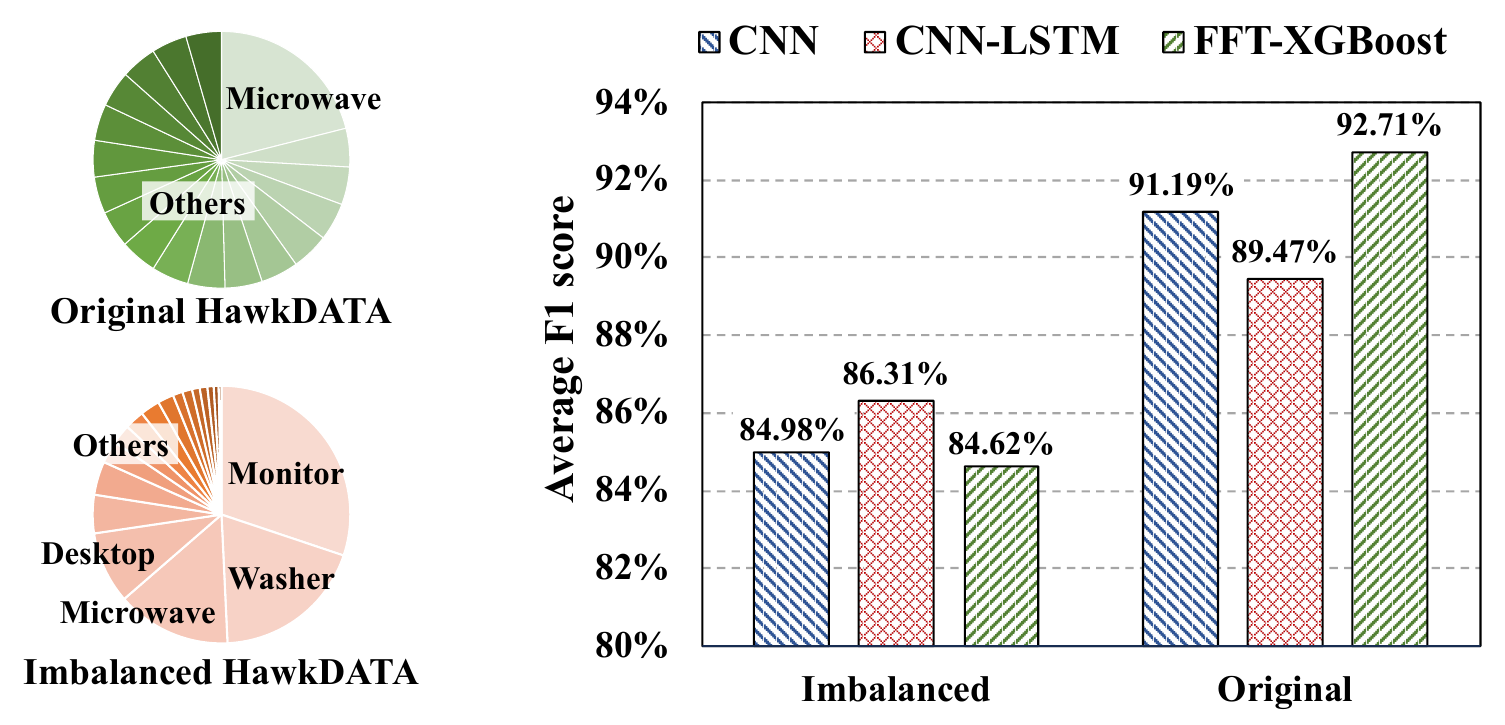}  
    \caption{Comparison of different classifier trained on different dataset.}
    \label{fig:compare-classifier-and-balance-effect}
  \end{minipage}
\end{figure}

Taking advantage of the enhanced SINR and robustness of the steady-state differential current algorithm pipeline, these basic network structures achieve high-accuracy event recognition, yielding similar final results. As shown in \autoref{fig:compare-classifier-and-balance-effect}, we choose FFT-XGBoost as our classifier, as it exhibited the highest recognition accuracy on the balanced dataset while maintaining minimal algorithmic overhead, enabling real-time recognition on embedded CPU platforms.

Regarding the impact of dataset balance, all classifiers demonstrated decreased recognition accuracy on the imbalanced dataset, even though most HawkDATA appliances are commonly used and the sample sizes are not significantly reduced. Additionally, different classifiers exhibited varying tolerances to dataset imbalance, with CNN-LSTM showing the least impact.

% \begin{table}[h]
% \centering
% \caption{Comparison of different classifier trained on different dataset.}
% \label{tab:compare-classifier-and-balance-effect}
% \scalebox{0.8}{
% \begin{tabular}{cccc}
% \hline
% \textbf{Model}      & \textbf{CNN} & \textbf{CNNLSTM} & \textbf{FFT-XGBoost} \\
% \hline
% \textbf{Imbalanced} & 84.98\% & 86.31\%  & 84.62\%  \\
% \textbf{Balanced}   & 91.19\% & 89.47\%  & \textbf{92.71\%} \\
% \hline
% \end{tabular}
% }
% \end{table}

\subsubsection{Impact of differential distance.}
\begin{figure}[t]
  \centering
    \includegraphics[width=0.9\linewidth]{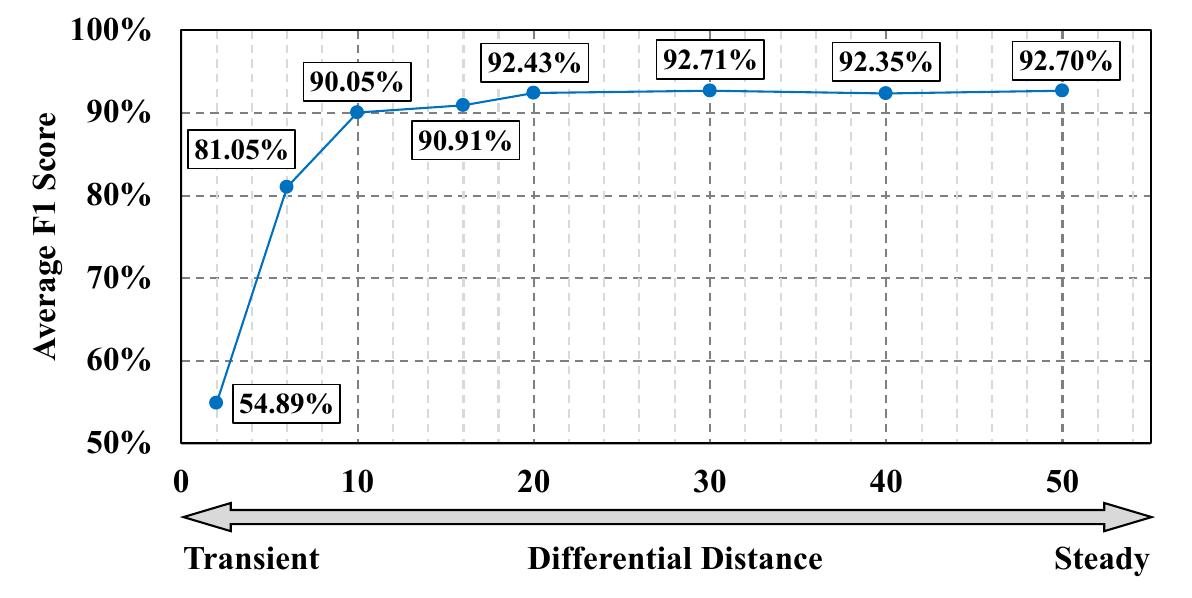}
    \caption{Impact of differential distance.}
    \label{fig:compare-diff-gap-effect}
\end{figure}

As illustrated in \autoref{fig:current-at-incandescent-lamp}, the distinction between steady and transient state differential current lies in the differential distance, which is consistent with the length of the voting window. A smaller differential distance results in a higher proportion of transient features within the voting window that cover events, while a larger differential distance increases the proportion of steady-state features. Moreover, the voting length affects the model's tolerance to false predictions; a larger differential interval increases the model's tolerance for errors. However, larger intervals also raise the likelihood of random noise interference and the occurrence of multiple events. Consequently, average accuracy exhibits fluctuations in \autoref{fig:compare-diff-gap-effect} when the interval is between 30 and 50. Ultimately, we select 30 to minimize the probability of multiple events occurring in the difference interval.

% As illustrated in \autoref{fig:current-at-incandescent-lamp}, the distinction between steady-state and transient-state differential current lies in the differential distance, which is consistent with the length of the voting window. A smaller differential distance results in a higher proportion of transient features within the voting window that cover events, while a larger differential distance increases the proportion of steady-state features. We explore different lengths of differential distances and train the optimal differential threshold on the training set, comparing the average F1 score of each appliance on the test set. Additionally, we constrain the interval to even lengths to mitigate the periodic change in the current of certain appliances, such as induction cookers. The final results are summarized in \autoref{fig:compare-diff-gap-effect}.

% \begin{table}[h]
%     \centering
%     \caption{Impact of differential distance}
%     \label{tab:compare-diff-gap-effect}
%     \scalebox{0.8}{
%     \begin{tabular}{cccccccc}
%     \hline
%         \textbf{DiffGap} & \textbf{2} & \textbf{6} & \textbf{10} & \textbf{20} & \textbf{30} & \textbf{40} & \textbf{50}  \\ \hline
%         Average & 54.89\% & 81.05\% & 90.05\% & 92.43\% & \textbf{92.71\%} & 92.35\% & 92.70\%  \\ \hline
%         % Average & 54.89\% & 81.05\% & 90.05\% & 90.91\% & 92.43\% & \textbf{92.71\%} & 92.35\% & 92.70\%  \\ \hline
%         % Average (>10w) & 57.79\% & 85.90\% & 96.27\% & 97.74\% & 98.51\% & 98.98\% & \textbf{99.05\%} & 99.02\%  \\ \hline
%     \end{tabular}
%     }
% \end{table}

% From the results in \autoref{fig:compare-diff-gap-effect}, as the differential interval increases from 2 to 30, the algorithm's robustness against false prediction improves, and recognition accuracy gradually increases, as discussed in section 4.1.2. However, accuracy declines from 30 to 50 due to increased fluctuation probability with longer differential distances. Ultimately, we selected 30 as the optimal differential distance for overall accuracy.

\subsubsection{State Identification Accuracy on HawkDATA}
\begin{figure}[t]
\centering
\includegraphics[width=0.9\linewidth]{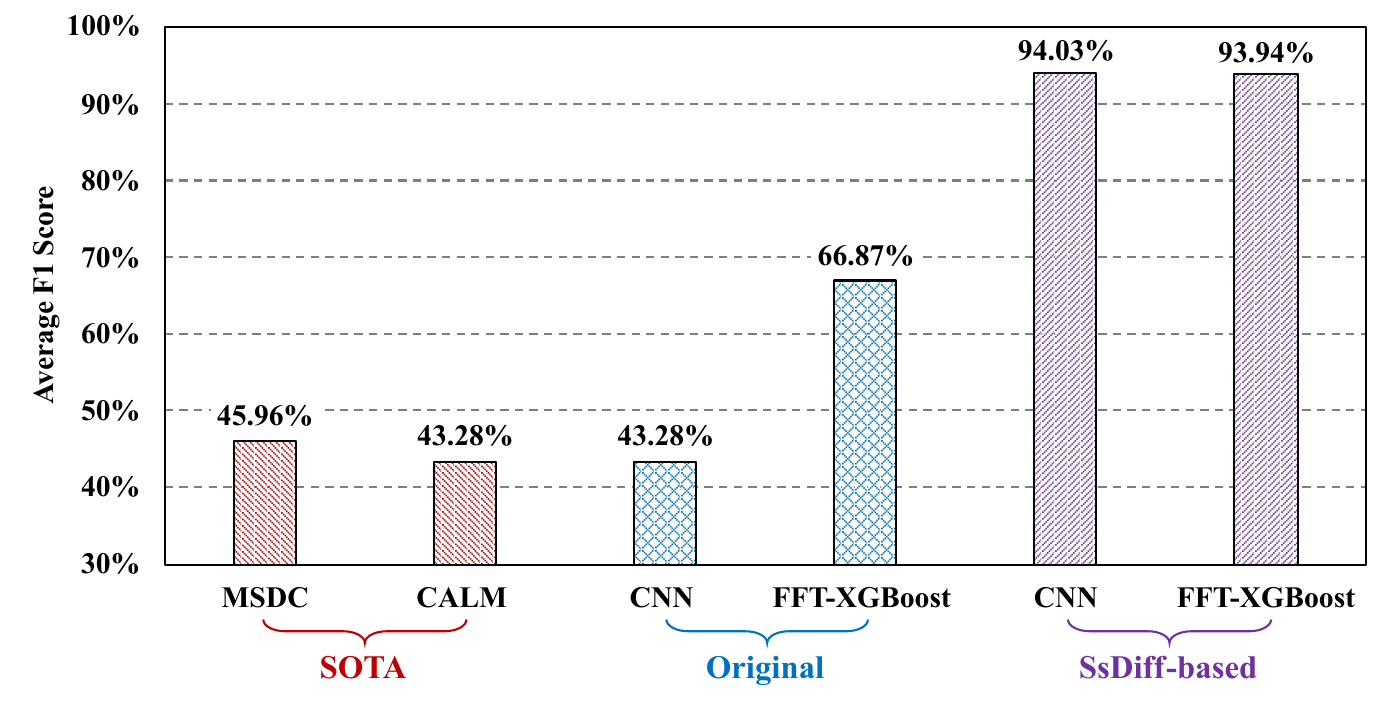}
    \caption{Accuracy comparison of state identification on HawkDATA. Original classifiers take raw data as input.}
    \label{fig:State-recognization-with-balanced-dataset}
\end{figure}

\autoref{fig:State-recognization-with-balanced-dataset} evaluates the impact of integrating steady-state differential current into the algorithm pipeline on state recognition accuracy and compares it with SOTA performance.

% \begin{table}[h]
%   \caption{Accuracy comparison of state identification on HawkDATA, classifiers with $^\dagger$ are intergrated in SsDiff-based algorithm pipeline.}
%   \label{tab:State-recognization-with-balanced-dataset}
%   \scalebox{0.7}{
%   \setlength{\tabcolsep}{1mm}{
%   \begin{tabular}{ccccccc}
%     \toprule    \textbf{Model}&\textbf{MSDC}\cite{he2023msdc}&\textbf{CALM}\cite{wang2021calm}&\textbf{CNN\cite{ismail2019deep}}&\textbf{FFT-XGBoost}&\textbf{CNN$^\dagger$}&\textbf{FFT-XGBoost$^\dagger$}\\
%     \midrule Average&45.96\%&43.28\%&43.28\%&66.87\%&\textbf{94.03\%}&93.94\%\\
%     % Average&-&-&-&-&93.54\%&93.53\%&92.28\%\\
%   \bottomrule
% \end{tabular}}
% }
% \end{table}

The results in \autoref{fig:State-recognization-with-balanced-dataset} show that the same XGBoost classifier, and CNNs with identical parameter sizes, significantly improve state recognition accuracy when integrating differential current processing.  Additionally, compared to the best SOTA performance, our simple classifier based on steady-state differential processing improved state recognition accuracy by 47.98\% to 48.07\%.

We observe that the previous two SOTA works performed poorly on HawkDATA. The evaluation results of the low-frequency signal-based MSDC align with earlier observations\cite{picon2016cooll, kahl2016whited}, behaving poorly in power trace decomposition. In contrast, the high-frequency signal-based CALM was evaluated on a single-appliance dataset.

% We also observe that MSDC has the highest identification accuracy on average among open-sourced works but falls far short of expectations which aligns to. CALM does not have the same accuracy as claimed in the paper (above 90\% for all appliances) because the validation datasets\cite{picon2016cooll, kahl2016whited} they use are collected from single-appliance currents instead of total currents from the main meter. The performance drop also reflects the difficulty of the NALM.

\subsubsection{Event Classification Accuracy on BLUED Dataset}

To validate the effectiveness of the Hawk algorithm pipeline for multi-state appliances and different AC standards, we trained the Hawk recognition model on the BLUED dataset for event classification. The event recognition of Hawk for multi-state appliances is similar to OFF-ON appliances, requiring pre-labeling all possible events for target appliances. The final recognition results are summarized in \autoref{tab:Event-recognization-with-BLUED}, compared with SOTA published results.

\begin{table}[h]
    \caption{Comparison of event classification accuracy on BLUED. Hawk is based on FFT-XGBoost.}
    \label{tab:Event-recognization-with-BLUED}
    \tabcolsep 2pt 
    \scalebox{0.8}{
    \begin{tabular}{ccccc}
        \toprule
            \textbf{Appliances} & \textbf{Power(w)} & \textbf{Number} & \textbf{MTShapelet\cite{yu2024multi}} & \textbf{Hawk}\\
        \midrule
            Hair Dryer & 1600 & 8 & 80.00\% & \textbf{100.00\%}\\
            Kitchen Aid Chopper & 1500 & 16 & 80.00\% & \textbf{93.75\%}\\
            Air Compressor & 1130 & 20 & 67.00\% & \textbf{100.00\%}\\
            Bedroom Lights & 190 & 19 & 86.00\% & \textbf{90.91\%}\\
            Fridge & 120 & 616 & 97.00\% & \textbf{99.67\%}\\
            Washroom light & 110 & 6 & \textbf{100.00\%} & \textbf{100.00\%}\\
            Bathroom upstairs lights & 65 & 98 & 85.00\% & \textbf{98.46\%}\\
            Backyard lights & 60 & 16 & 89.00\% & \textbf{93.75\%}\\
        \midrule
            Average & -& -& 85.50\%& \textbf{97.07\%}\\
            Weighted average & -& -& 95.00\%& \textbf{99.13\%}\\
        \bottomrule
    \end{tabular}}
\end{table}
According to \autoref{tab:Event-recognization-with-BLUED}, Hawk event classification average F1 score outperforms MTS-Shapelet by 11.57\%, and the weighted average F1 score improved by 4.13\%.
Moreover, higher recognition accuracy is achieved across all appliances, with three appliances achieving F1 scores of 100\%. MTS-Shapelet and Hawk achieve 100\% classification accuracy on the Washroom light appliance, which only has six samples in total, indicating a potential issue of insufficient validation and highlighting the importance of data balance.

\subsection{In-The-Wild Evaluation}
\subsubsection{Overall Setup}
In the field experiments, our objective is to validate the effectiveness of the Hawk system in real world scenarios, which includes computational overhead, real-time performance, and recognition accuracy in various unknown background appliance settings. The Hawk model was deployed without modification in two real scenarios that align with the application scenario.

\textit{Computational Platforms.} To validate Hawk run-time performance, we performed real-time streaming inference performance assessments on three CPU-based platforms, as listed in \autoref{tab:runtime-on-platform}. 
% The three typical platforms correspond to three possible scenarios: NALM on the PC (DELL-Precision 3440) as a home or industrial application, NALM on edge servers (NVIDIA Jetson Orin) to enable green communities, and NALM on an embedded PC (Rasoberry PI) as a smart meter.

\textit{Metrics.} We use the F1 score to evaluate the accuracy of Hawk system recognition, with the average F1 score used to assess overall performance. For runtime performance, we measure the Hawk's average streaming inference latency and the actual memory footprint (Residential Set Size, RSS) of the entire program. 

\subsubsection{Run-time Performance Evaluation}
This section aims to evaluate the performance of streaming inference, which processes individual samples sequentially. The real-time requirement for streaming inference is that the average inference latency must be less than the average generation delay (20 ms in our AC settings). This is a stringent performance criterion and are proved valuable in real application scenarios\cite{armel2013disaggregation}. Performance on the three platforms is summarized in \autoref{tab:runtime-on-platform}.

\begin{table}[h]
  \caption{Average streaming inference latency and memory footprint of Hawk on three distinct platforms.}
  \label{tab:runtime-on-platform}
  \tabcolsep 0.5pt 
  \scalebox{0.8}{
  \begin{tabular}{ccccc}
    \toprule
    \textbf{Platform}&\textbf{CPU}&\textbf{Latency (ms)} & \hspace{0.2cm} \textbf{Mem (MB)} \\
    \midrule
    Desktop&Intel i7-10700@2.9GHz& 3.31 & 42.50\\
    Jetson Orin NX&Cortex-A78AE@2GHz&4.07&40.19\\
    Raspberry PI 4B&Cortex-A72@1.8GHz& 6.59 & 32.60 \\
    \midrule
    Real-time Requirement & Single core & $<$ 20 & -\\
  \bottomrule
  \end{tabular}
  }
\end{table}

\autoref{tab:runtime-on-platform} shows that our model satisfies the real-time requirement of immediate inference upon data generation on all three platforms. For future application developers, we recommend using buffered data for batch processing, parallel programming, and filter samples based on differential power for performance optimization.
% These approaches have proven to be effective in our engineering exploration.

\subsubsection{Event Recognition Accuracy Evaluation}
The event recognition accuracy in two real environments is summarized in \autoref{tab:in-the-wild-accuracy}.

\begin{table}[h]
  \caption{Event recognition accuracy in two real settings.}
  % \caption{Event recognition accuracy in two in-the-wild settings. The accuracy of the HawkDATA is compared.}
  \label{tab:in-the-wild-accuracy}
  \scalebox{0.8}{
  \begin{tabular}{ccccc}
    \toprule
    \textbf{Appliances}&\textbf{Power(w)}&\textbf{HawkDATA}&\textbf{Residence}&\textbf{Office} \\
    \midrule
        AirHeater  & 2160 & \textbf{100.00\%}    & -  &  98.00\%   \\
        ElectricKettle & 1500 & \textbf{100.00\%} & 96.29\%   & -      \\
        MicrowaveOven  & 1315 & \textbf{100.00\%}  & 98.82\%& -      \\
        ElectricCooker & 500 & \textbf{99.60\%} & 98.71\% & -      \\
        Desktop  & 40 & \textbf{99.18\%} & -  &  93.19\% \\
        Humidifier & 40  & \textbf{100.00\%}   & -  & 85.41\% \\
        LEDLamp36W & 36 & \textbf{99.58\%} & 94.88\%   &  98.81\%\\
        LEDLamp24W    & 24 & 96.52\% & 91.42\%   & \textbf{96.83}\% \\
        FluorescentLamp  & 19 & 96.20\%    & -  &  \textbf{96.30}\% \\
    \midrule
        Average & - &  \textbf{99.01}\% &  96.02\% &   94.76\% \\
  \bottomrule
  \end{tabular}
  }
\end{table}
As shown in \autoref{tab:in-the-wild-accuracy}, all appliances except for the humidifier experienced less than a 10\% decline in F1 score compared with accuracy on HawkDATA. The recognition accuracy of some low-power appliances is even improved in office environments.
This experiment meets our initial goals and demonstrates the effectiveness of the Hawk system. Notably, even with many continuously changing background appliances, such as laptops and desktops, low-power appliances are still accurately recognized. However, the accuracy drop for the humidifier is significant. We analyze the reasons for this performance decline in the next section.

\section{Discussion And Future work} \label{discussion}
In this section, we discuss some limitations of the Hawk system and analysis of recognition failures.

\textit{Limitations of Binary Modeling}: The balanced Gray code-based strategy models appliance OFF-ON states as binary bit value, which does not accommodate multi-state appliances. 
OFF-ON appliances can be considered a particular case of multi-state appliances. The performance of Hawk's algorithm pipeline on the BLUED dataset, which includes multi-state appliances, proves that the processes of recognizing OFF-ON events and multi-state events are interconnected. However, this characteristic limits the diversity of the current HawkDATA appliance collection. A new balanced abstraction to model multi-state appliances is needed.

\textit{Limitations of Application Scenarios}: Recognizing appliances based on their model rather than their type is more practical but more restrictive. The current Hawk model still requires data collection from specific appliance sets, which remains a significant overhead. Future work could reduce this overhead through multimodal unsupervised approaches \cite{zhu2023combining} and few-shot learning, and our proposed algorithm pipeline design will benefit these efforts.

\textit{Limitations of Appliance Selection}: Despite careful selection, the number of appliances in the HawkDATA collection is still limited compared to the variety of background appliances. For instance, the humidifier with the highest-power LC circuit suffers numerous false positive results in field experiments. Expanding the variety of appliance types and usage periods will be a future effort.

\textit{Failures Due to Voltage Variations}:
Voltage fluctuations in residential areas contribute to most false-negative results of electric kettle recognition. In real-world environments, the voltage of the appliances fluctuates over time and shows different trends in different areas. Moreover, different types of appliances exhibit different current responses to the changes: resistive appliances show a positive correlation between current and voltage, while capacitive appliances (e.g., LED light) with AC-DC front ends show a negative correlation to keep output power stable. Future work will further analyze the impact of voltage on recognition results and incorporate voltage waveforms for more precise appliance identification.

\textit{Falures Due to Complex Appliances}:
Hawk achieves its goals and performs well with low-power appliance events recognition in selected two real scenarios. However, the algorithm pipeline still struggles with recognizing ultra-low power (<10w), continuously varying power, and among identical appliances. 
Future work will focus on reducing background noise interference and distinguishing between different appliances of the same model\cite{gupta2023enabling}.

\section{CONCLUSION} \label{conclusion}
Non-intrusive appliance load monitoring(NALM) faces the challenges of high dataset construction overhead and low appliance identification accuracy. This paper presents Hawk, an efficient NALM system for accurate low-power appliance recognition. To improve the efficiency of dataset construction, we propose an efficient dataset construction scheme based on balanced Gray codes and a sampling synchronization strategy based on shared perceptible time to enable automatic data annotation.
Taking advantage of the enhanced SINR and robustness provided by the steady-state differential current-based algorithm pipeline, some basic classifiers perform high accuracy and low computational overhead in real-time recognition across different tasks, even with low-power appliances.
To our knowledge, Hawk is the first NALM system to accurately and in real time identify low-power appliance usages in real-world scenarios, with an average cost of stable high-frequency current sampling of only \$4.12. A balanced and diverse NALM dataset, HawkDATA, has been published.

\section{Acknowledgement}
We are grateful to anonymous reviewers for their constructive suggestions. We sincerely thank Prof. Xingwu Liu, PhD candidates Shun Lu and Zheming Yang, and Dr. Xiaoyu Wang for their valuable suggestions on this paper. We also appreciate the Nanjing Institute of InfoSuperBahn for providing the data collection environment. This research was partly supported by the National Natural ScienceFoundation of China under Grant Nos. 62402475 and 62072434, and the Innovation Funding of ICT, CAS under Grant No. E361050.
% dasfdfadsfasdfasdfasdf
\newpage
\balance
\bibliographystyle{ACM-Reference-Format}
\bibliography{sample-base}

% \appendix
% \newpage
% \section{Artifact Evaluation}
% \subsection{Overview}:
% We will perform an AE on the results of our FFT-XGBoost implementation, which presents our best experimental results. 
% \subsection{Project Location:}

% The following is the location of the project.

% Our code has been published anonymously and is open source at \url{https://anonymous.4open.science/r/Hawk4NALM2/}. This link includes the download for our HawkDATA as well as our code. The software requirements and validation process are outlined in the README on GitHub.

\end{document}